\documentclass[manuscript]{acmart}

\AtBeginDocument{%
  }

\setcopyright{acmcopyright}
\copyrightyear{2023}
\acmYear{2023}
\acmDOI{XXXXXXX.XXXXXXX}

\usepackage{tabularx}
\usepackage{bbding}
\usepackage{wasysym}
\usepackage{threeparttable}
\usepackage{stfloats}
\usepackage{multirow}
\usepackage{makecell}
\usepackage{color}
\usepackage{enumerate}
\usepackage{listings}
\usepackage{xcolor}
\usepackage{adjustbox}
\usepackage{enumitem}

\acmJournal{CSUR}
\acmVolume{37}
\acmNumber{4}
\acmArticle{47}
\acmMonth{6}

\lstdefinelanguage{Solidity}{
	keywords=[1]{anonymous, assembly, assert, balance, break, call, callcode, case, catch, class, constant, continue, constructor, contract, debugger, default, delegatecall, delete, do, else, emit, event, experimental, export, external, false, finally, for, function, gas, if, implements, import, in, indexed, instanceof, interface, internal, is, length, library, log0, log1, log2, log3, log4, memory, modifier, new, payable, pragma, private, protected, public, pure, push, require, return, returns, revert, selfdestruct, send, solidity, storage, struct, suicide, super, switch, then, this, throw, transfer, true, try, typeof, using, value, view, while, with, addmod, ecrecover, keccak256, mulmod, ripemd160, sha256, sha3}, 
	keywordstyle=[1]\color{blue}\bfseries,
	keywords=[2]{address, bool, byte, bytes, bytes1, bytes2, bytes3, bytes4, bytes5, bytes6, bytes7, bytes8, bytes9, bytes10, bytes11, bytes12, bytes13, bytes14, bytes15, bytes16, bytes17, bytes18, bytes19, bytes20, bytes21, bytes22, bytes23, bytes24, bytes25, bytes26, bytes27, bytes28, bytes29, bytes30, bytes31, bytes32, enum, int, int8, int16, int24, int32, int40, int48, int56, int64, int72, int80, int88, int96, int104, int112, int120, int128, int136, int144, int152, int160, int168, int176, int184, int192, int200, int208, int216, int224, int232, int240, int248, int256, mapping, string, uint, uint8, uint16, uint24, uint32, uint40, uint48, uint56, uint64, uint72, uint80, uint88, uint96, uint104, uint112, uint120, uint128, uint136, uint144, uint152, uint160, uint168, uint176, uint184, uint192, uint200, uint208, uint216, uint224, uint232, uint240, uint248, uint256, var, void, ether, finney, szabo, wei, days, hours, minutes, seconds, weeks, years},	
	keywordstyle=[2]\color{teal}\bfseries,
	keywords=[3]{block, blockhash, coinbase, difficulty, gaslimit, number, timestamp, msg, data, gas, sender, sig, value, now, tx, gasprice, origin},	
	keywordstyle=[3]\color{violet}\bfseries,
	identifierstyle=\color{black},
	sensitive=true,
	comment=[l]{//},
	morecomment=[s]{/*}{*/},
	commentstyle=\color{gray}\ttfamily,
	stringstyle=\color{red}\ttfamily,
	morestring=[b]',
	morestring=[b]"
}

\begin{document}

\title{Survey on Quality Assurance of Smart Contracts }



\author{Zhiyuan Wei}
\affiliation{%
  \institution{Beijing Institute of Technology}
  \city{Beijing}
  \country{China}
}
\email{weizhiyuan@bit.edu.cn}

\author{Jing Sun}
\affiliation{%
  \institution{University of Auckland}
  \city{Auckland}
  \country{New Zealand}
}
\email{jing.sun@auckland.ac.nz}

\author{Zijian Zhang}
\authornotemark[1]
\email{zhangzijian@bit.edu.cn}
\author{Xianhao Zhang}
\email{1120191274@bit.edu.cn}
\author{Xiaoxuan Yang}
\email{yangxiaoxuan979@gmail.com}
\author{Liehuang Zhu}
\email{liehuangz@bit.edu.cn}
\affiliation{%
  \institution{Beijing Institute of Technology}
  \city{Beijing}
  \country{China}
}

\renewcommand{\shortauthors}{Wei et al.}

\begin{abstract}
As blockchain technology continues to advance, the secure deployment of smart contracts has become increasingly prevalent, underscoring the critical need for robust security measures. This surge in usage has led to a rise in security breaches, often resulting in substantial financial losses for users. This paper presents a comprehensive survey of smart contract quality assurance, from understanding vulnerabilities to evaluating the effectiveness of detection tools. 
Our work is notable for its innovative classification of forty smart contract vulnerabilities, mapping them to established attack patterns. We further examine nine defense mechanisms, assessing their efficacy in mitigating smart contract attacks. Furthermore, we develop a labeled dataset as a benchmark encompassing ten common vulnerability types, which serves as a critical resource for future research. We also conduct comprehensive experiments to evaluate fourteen vulnerability detection tools, providing a comparative analysis that highlights their strengths and limitations. In summary, this survey synthesizes state-of-the-art knowledge in smart contract security, offering practical recommendations to guide future research and foster the development of robust security practices in the field.
\end{abstract}



\received{31 Jun 2024}
\received[revised]{10 Aug 2024}
\received[accepted]{5 September 2024}

\maketitle

\section{Introduction}
Blockchain technology has gained prominence in academia and industry as a secure and private solution for diverse applications \cite{wood2014ethereum, hughes2019blockchain}. As a distributed ledger, blockchain technology is replicated and shared among a network of peer-to-peer nodes. It eliminates the need for intermediaries, thereby providing decentralization, transparency, immutability, security, and reliability. By maintaining a chronologically growing and immutable data record, blockchain systems have become ideal for a multitude of domains, ranging from business to healthcare. 
In the realm of business, blockchain significantly reduces operational costs \cite{casino2019systematic}. Healthcare applications of blockchain range from securing cloud-based cyber-physical systems \cite{gupta2021blockchain, nguyen2021secure} to enhancing privacy in monitoring systems \cite{raj2022privacy}. Its role in managing healthcare challenges, particularly during the COVID-19 pandemic, further exemplifies its broad utility and adaptability \cite{martinez2022analysis}. Additionally, blockchain's integration with AI in areas like resource allocation for drone-terrestrial networks indicates its expanding scope in various innovative fields \cite{pan2022blockchain}.

Smart contracts have emerged as a pivotal component of blockchain technology.  Originally conceptualized in the early 1990s, they gained significant traction with the advent of blockchain platforms like Ethereum. These self-executing contracts, implemented through lines of code \cite{LuuCOSH16, AtzeiBC16}, have transformed traditional contract enforcement by automating it and embedding it within the blockchain network. They ensure transparency and immutability of contract rules, enforced by blockchain network participants. They allow decentralized applications (DApps) to be built on the blockchain networks, facilitating applications in diverse domains such as financial services \cite{ZhangWLSL22}, healthcare \cite{raj2022privacy, QahtanSZOZA23}, the Internet of Things \cite{GuptaSK23}, crowdfunding \cite{SaadatRNZ19}, and supply chain management \cite{SinghMGM23, SultanaTK22}. 

Typically, smart contracts are associated with the native cryptocurrency of the blockchain, which is used to compensate network participants for contract execution. This incentivization maintains the blockchain's security and decentralization. The novelty and potential impact of smart contracts in revolutionizing business transactions have attracted significant research interest. However, despite growing attention, numerous open research questions remain to be addressed in this emerging field.

Compared to traditional programs, smart contracts possess unique characteristics that make them more vulnerable to software attacks. Firstly, smart contracts are immutable, meaning that once deployed on the blockchain, their code cannot be modified. While immutability ensures trust and transparency, it also means that any vulnerabilities or errors in the code cannot be easily rectified without deploying a new contract version. Secondly, smart contracts frequently manage valuable digital assets, such as cryptocurrencies or digital tokens. This concentration of value assets attracts malicious users who actively seek to exploit vulnerabilities in the contract code. Thirdly, public blockchains, such as Ethereum, are permissionless, allowing universal access and interaction with deployed smart contracts. This open nature lowers entry barrier for potential attackers, making it easier for them to identify and exploit vulnerabilities in smart contracts. Moreover, the common practice of publishing contract code on platforms such as Etherscan further exposes it to potential security threats.

The exploitation of these vulnerabilities has led to significant financial losses and broader implications in the blockchain world. For instance, the 2016 DAO attack, where attackers diverted over 3.6 million ether, resulted in a loss of around 70 million USD and triggered a substantial drop in ether's value \cite{zhou2023sok}. The Parity Multisig Wallet incident led to a loss of about 30 million USD \cite{Palladino2017}. In 2020, the KuCoin hack resulted in the loss of over 280 million USD worth of cryptocurrencies. This incident underscored the risks inherent in smart contracts and catalyzed discussions on enhancing security measures and potential regulatory frameworks. Furthermore, the February 2020 reentrancy attack on the bZx DeFi platform, resulting in losses of nearly 350,000 USD, highlighted vulnerabilities in complex DeFi smart contracts and emphasized the necessity for rigorous testing and security audits \cite{timcopeland2020}.
Beyond the immediate financial losses, these incidents have raised concerns about the security and reliability of smart contracts, impacting user trust and investor confidence. 

\begin{table}[t]
\centering
\footnotesize
\setlength{\tabcolsep}{4pt}
\caption{Comparison of Surveys on Smart Contract Security}
\label{tab:comparison_survey}
\begin{tabular}{@{}lcccccc@{}}
\toprule
\textbf{Research} & \textbf{Venue}& \textbf{Vulnerability} & \textbf{Attacks} & \textbf{Defense} & \textbf{Tools} & \textbf{Datasets} \\
 & & \textbf{Types} & & \textbf{Methods} &  & \\
\midrule
Atzei et al. \cite{AtzeiBC16} & ETAPS'17 &12 & 9 & 3 & -- & -- \\
Zheng et al. \cite{zheng2020overview} & FGCS Journal'20&\checkmark & -- & -- & -- & -- \\
Chen et al. \cite{ChenPNX20} & ACM Comput. Surv. &\textbf{40} & \textbf{29} & 8 & -- & -- \\
Angelo et al. \cite{AngeloS19} & DAPPCON'19 &-- & -- & -- & 27 & -- \\
Durieux et al. \cite{DurieuxFAC20} &  ICSE’20 &10 & -- & -- & 35 & \checkmark \\
Tolmach et al. \cite{TolmachLLLL22} & ACM Comput. Surv. & -- & -- & 5 & 34 & -- \\
Ivanov et al. \cite{ivanov2023security}& ACM Comput. Surv. & 37 & -- & 8 & 38 & -- \\
Chu et al. \cite{chu2023survey} & IST Journal'23 & 12 & -- & 2 & 20 & \checkmark  \\
He et al. \cite{he2023detection}& IEEE IoT Journal'23 & 8 & -- & 6 & 6 & -- \\
Zhang et al. \cite{zhang2023demystifying} & ICSE'23 & 14 & -- & 6 & 5 & \checkmark \\
Chaliasos et al. \cite{chaliasos2024smart} & ICSE'24 & 14 & -- & 6 & 5 & \checkmark  \\
This work & -- &\textbf{40}  & 9 & \textbf{9} & \textbf{101} & \checkmark \\
\bottomrule
\end{tabular}
\end{table}

Due to the novelty and potential impact of smart contracts, there are some notable surveys involving vulnerable smart contracts from various perspectives. Table \ref{tab:comparison_survey} underscores the alignment and distinctions between our work and previous surveys. Our study specifically addresses various aspects of smart contract security, including vulnerability types, attacks, defense methods, detection technologies, and relevant datasets.

Atzeri et al. \cite{AtzeiBC16} pioneered the survey of smart contract security, categorizing 12 vulnerabilities into three domains: Solidity, EVM bytecode, and blockchain. This classification framework has been widely adopted in subsequent research and has practical applications in identifying and mitigating security risks in real-world smart contract deployments. Zheng et al. \cite{zheng2020overview}  conducted a comparative analysis of various smart contract platforms and developed a taxonomy of applications. Their study evaluated the features and characteristics of diverse platforms, emphasizing the potential impact of vulnerabilities on sectors such as finance and healthcare, where transaction integrity and security are paramount.

Chen et al. \cite{ChenPNX20} expanded the scope beyond vulnerability analysis to encompass defense mechanisms for blockchain security. Their research examined various defense techniques and strategies employed to bolster smart contract security. Angelo and Salzer \cite{AngeloS19} conducted a comprehensive survey on vulnerability detection tools tailored for Ethereum smart contracts. Their study encompassed tools from both academia and industry, offering insights into the array of available vulnerability identification tools. Durieux et al. \cite{DurieuxFAC20} performed a comprehensive evaluation of nine smart contract detection tools. They assessed these tools using a labeled dataset and numerous real-world smart contracts, yielding a thorough analysis of their efficacy.

Tolmach et al. \cite{TolmachLLLL22} concentrated on the formal verification of smart contracts across diverse applications. Their research explored the application of formal verification techniques for ensuring the correctness and security of smart contracts. Ivanov et al. \cite{ivanov2023security} focused on security threat mitigation, developing a concise vulnerability map encompassing 37 known vulnerabilities. This map synthesized the vulnerability-addressing capabilities of 38 distinct classes of threat mitigation solutions. Chu et al. \cite{chu2023survey} examined vulnerability data sources, detection methods, and repair techniques. Through automated vulnerability injection into contracts, they constructed an objective vulnerability dataset, enhancing the analysis of existing security methods' performance.


He et al. \cite{he2023detection} evaluated performance detection methods and their associated tools. They primarily tested six detection tools, analyzing them from multiple perspectives: detection accuracy, execution time, and Solidity version compatibility.
Zhang et al. \cite{zhang2023demystifying} conducted a systematic investigation of 462 defects reported in CodeArena audits and 54 exploits, assessing the detection capabilities of existing tools. Their survey indicates that existing tools can detect machine-auditable vulnerabilities, with more than 80\% of exploitable bugs falling into this category. Some vulnerabilities are too complex or subtle and require the expertise of multiple human auditors. 
Chaliasos et al. \cite{chaliasos2024smart} compared five state-of-the-art (SOTA) automated security tools with 49 developers and auditors from leading DeFi protocols. Their findings revealed that the tools could have prevented only 8\% of the attacks in their dataset.


Although previous surveys offer valuable insights into specific aspects of smart contract security, they often lack a comprehensive analysis encompassing all perspectives of vulnerable smart contracts. It is crucial to develop a thorough understanding of vulnerabilities, attacks, defenses, and tool evaluation to gain a holistic view of the challenges and potential solutions related to smart contract security. Our paper aims to bridge this gap by incorporating multiple perspectives, offering a more comprehensive analysis. Through a systematic examination of vulnerabilities, attacks, defenses, and tools, we strive to provide a comprehensive understanding of the challenges posed by vulnerable smart contracts and explore viable solutions.

The primary objective of our paper is to contribute to the existing body of research on smart contract security by providing a comprehensive and up-to-date analysis. To achieve this objective, we present the following key contributions:

\begin{itemize}[leftmargin=*] 
  \item{\textbf{Novel Vulnerability Classification}:} We propose a novel vulnerability classification that enhances the understanding of the underlying causes of vulnerabilities in smart contracts. This classification facilitates more effective categorization and analysis of vulnerabilities by researchers, establishing a robust foundation for security measures and enhanced vulnerability management.
  \item{\textbf{In-depth Analysis of Real-World Attacks}:} We present a comprehensive analysis of real-world attacks on smart contracts to gain valuable insights into the methods employed by attackers and the potential consequences of these attacks. By examining and dissecting these attacks, we aim to provide a clearer understanding of the exploitation of vulnerabilities in practice, empowering developers and auditors to proactively address potential threats.
  \item{\textbf{Exploration of Defense Mechanisms}:} We conduct a rigorous assessment of existing defense mechanisms employed to mitigate smart contract attacks. Through this exploration, we identify areas for improvement and potential new approaches to enhance the security of smart contracts. By analyzing the strengths and weaknesses of current defense mechanisms, we aim to contribute to the development of more robust and effective security practices.
  \item{\textbf{Evaluation of Vulnerability-Detecting Tools}:} We conduct a comprehensive evaluation of 14 representative vulnerability-detecting tools used in smart contract analysis. This evaluation encompasses the accuracy, performance, and effectiveness of each tool. By providing insights into the strengths and weaknesses of these tools, we assist researchers and practitioners in selecting the most suitable tools for identifying vulnerabilities in smart contracts.
  \item{\textbf{Benchmark Dataset for Tool Evaluation}:}  We introduce a comprehensive benchmark dataset, encompassing 100 vulnerable smart contract cases and 10 secure contract instances, to facilitate rigorous evaluation of vulnerability detection tools. This meticulously curated dataset serves as a standardized reference point for assessing the efficacy of vulnerability detection tools. It enables fair and objective comparisons, allowing researchers and practitioners to assess the capabilities of different tools in a consistent manner.
\end{itemize}

Through these key contributions, our paper aims to serve as a comprehensive resource for researchers, developers, and auditors in the field of smart contract security. We aspire to advance secure smart contract development practices and promote the widespread adoption of secure smart contracts in real-world applications.

The rest of this paper is constructed as follows. Section \ref{Overview} provides a brief overview of smart contract platforms and discusses the methodology employed in this survey to ensure a comprehensive analysis. Section \ref{vulnerability} presents a detailed analysis of 40 smart contract vulnerabilities, examining their root causes and shedding light on the underlying factors that contribute to their existence. Section \ref{attacks} explores 8 representative attacks and demonstrates how these vulnerabilities can be exploited. By illustrating real-world attack scenarios, we aim to enhance our understanding of the potential consequences of these vulnerabilities. Section \ref{defense} examines various defense methodologies and repair techniques available for smart contracts. We discuss the effectiveness of these defensive measures in mitigating vulnerabilities and reducing the likelihood of successful attacks. Section \ref{evaluation} focuses on the evaluation of 14 commonly used tools for detecting vulnerabilities in smart contracts. We assess the accuracy and performance of these tools by subjecting them to a rigorous evaluation against a carefully curated benchmarking dataset. Section \ref{discussion} concludes the paper by summarizing the key findings and contributions. We also discuss future directions and potential research areas to further advance the field of smart contract security. 

\section{Overview of Smart Contracts and Survey Methodology}
\label{Overview}
\subsection{Smart Contracts}

The foundational concept of smart contracts is attributed to Nick Szabo, who pioneered the idea in 1996, predating the creation of blockchain technology. Szabo envisioned smart contracts as computer protocols that enable parties to engage in digitally verifiable and self-executing agreements. These contracts are written in code format and facilitate secure and efficient transactions, reducing costs and expediting execution compared to traditional contracts \cite{Szabo2018SmartC}. However, the development of smart contracts faced challenges in trustless systems until the emergence of Ethereum in 2015 \cite{KhanLGBB21}. Ethereum's introduction of a blockchain-based platform specifically designed for executing smart contracts revolutionized the field. Ethereum's success paved the way for the proliferation of other blockchain platforms that also support smart contract development. These platforms include Hyperledger \cite{Saad2020}, EOSIO \cite{HeZ00L0YJ21}, Tezos \cite{BernardoCCJPT20}, NEO \cite{NguyenDT19}, and even Bitcoin, which implemented its own version of smart contracts \cite{BartolettiZ19}. 

Our survey presents a comprehensive comparative analysis of four prominent blockchain platforms: Bition, Ethereum, Hyperledger Fabric, and EOS. Table \ref{tab:smart_contracts} presents this comparison, highlighting the similarities and differences in terms of smart contract implementation and security across these diverse platforms. Our survey specifically targets these platforms due to their widespread adoption and distinct approaches to handling smart contracts. Bitcoin, although not primarily known for complex smart contracts, is included for its foundational role in the blockchain space and its recent developments in enabling basic smart contract functionalities. Ethereum is renowned for its pioneering role in smart contract technology, emphasizing decentralization and developer accessibility. Hyperledger Fabric offers a modular and customizable architecture, making it more suited to enterprise applications. EOS, while bearing some similarities to Ethereum, differentiates itself with unique governance models and performance optimizations. 

\begin{table}[t]
\centering
\footnotesize
\setlength{\tabcolsep}{4pt}
\caption{Comparative Analysis of Smart Contract Ecosystems Across Major Blockchain Platforms}
\label{tab:smart_contracts}
\begin{tabularx}{\textwidth}{@{}l*{4}{>{\raggedright\arraybackslash}X}@{}}
\toprule
\textbf{Feature} & \textbf{Bitcoin} & \textbf{Ethereum} & \textbf{Hyperledger Fabric} & \textbf{EOSIO} \\
\midrule
Language & Script & Solidity & Go, JavaScript, Java & C++, Python, JavaScript \\ \hline
Consensus & PoW & PoW/PoS & PBFT/Raft & DPoS+BFT \\ \hline
Environment & Script-based & EVM & Docker & EOS VM \\ \hline
Cryptocurrency & Bitcoin (BTC) & Ether (ETH) & None & EOS \\ \hline
Scalability & High & High & Low & High \\ \hline
Speed (TPS) & 27 & 30 & -- & 4,000 \\ \hline
Confirmation & >1000s & <100s & <10s & <10s \\ \hline
Applications & Send/receive messages, verify signatures, transactions & Finance, supply chain, IoT, medical & Enterprise licensing, finance & Finance, gambling \\ \hline
Security Issues & Informal cryptographic protocols, poorly documented features & Source code/EVM errors, blockchain attacks & Source code errors, blockchain attacks & Source code errors, transaction verification lacks \\ \hline
Solutions & Secure cryptographic protocol design & Vulnerability detection tools, contract monitoring & Vulnerability detection tools, upgrades & Vulnerability detection tools, upgrades \\ \hline
Related Survey & \cite{BartolettiZ18a,BartolettiLMZ22,AtzeiBCLZ18} & \cite{AtzeiBC17,AngeloS19,ChenPNX20, zhou2020ever} & \cite{Saad2020,VaccaSVC21} & \cite{abs-2207-09227} \\ 
\bottomrule
\end{tabularx}
\end{table}

The following sections analyze four blockchain platforms, highlighting their unique characteristics and approaches to smart contract technology.

\begin{itemize}[leftmargin=*] 
  \item \textbf{Bitcoin}, the first blockchain system, primarily facilitate secure peer-to-peer currency transactions without the need for intermediaries. Although not originally designed to support extensive smart contract functionalities like Ethereum, Bitcoin offers limited capabilities for executing cryptographic protocols. These capabilities include basic operations such as sending/receiving messages, verifying signatures, and searching transactions.
  Bitcoin smart contracts, often called script-based contracts, are primarily written in Script, a simple, stack-based language designed specifically for transactions.  These smart contracts on Bitcoin are often referred to as script-based contracts. Although Bitcoin smart contracts are relatively simple and have limited functionality, they still serve as a valuable tool for executing basic transactional functions in a trustless and decentralized manner. Researchers have studied Bitcoin's scripting language construction and capabilities, with notable examples including Ivy, BALZAC, SIMPLICIT, and BITML \cite{BartolettiZ19}. These studies explored different aspects of Bitcoin's scripting language and its applications in cryptographic protocols.

  \item \textbf{Ehtereum} pioneered the deployment of smart contracts on blockchain platforms. It enables developers to create their own smart contracts using Solidity, a Turing-complete programming language. These smart contracts are executed on the Ethereum Virtual Machine (EVM), which provides a runtime environment for their execution. Ethereum's smart contracts have significantly expanded blockchain technology applications, giving rise to Decentralized Applications (DApps) that enable trustless, decentralized interactions. Ethereum adopts an account-centered model for smart contracts, contrasting with Bitcoin's Unspent Transaction Output (UTXO) model. Ethereum's success with smart contracts has established it as a prominent second-generation blockchain platform, often referred to as blockchain 2.0 \cite{ChenPNX20}. Solidity is the most widely used programming language for Ethereum smart contracts on Ethereum, is specifically designed to compile into EVM bytecode, the low-level representation of smart contracts.
  
  \item \textbf{Hyperledger Fabric (HF)} is an enterprise-focused blockchain platform operating on a permissioned network. Unlike public blockchains, HF allows participation of trusted organizations through a membership service provider \cite{AndroulakiBBCCC18}. This approach suits scenarios involving collaboration among multiple business organizations, offering enhanced privacy, scalability, and network control.
  HF smart contracts, called \textit{chaincode}, offer flexibility in programming language support. Developers can choose from various programming languages such as Golang, Java, and JavaScript to implement their smart contracts, allowing them to leverage their existing expertise and use the language most suitable for their application \cite{KhanLGBB21}. Privacy is another key feature of HF smart contracts. Hyperledger Fabric supports private transactions, enabling selected participants to engage in confidential transactions without revealing sensitive information to all network participants. HF smart contracts also feature modular design, facilitating easier maintenance and updates \cite{LiLDYZZL22}.
  
  \item \textbf{EOSIO} introduces innovative features to the field of blockchain technology, e.g., delegated proof of stake consensus (DPOS) and updatable smart contracts \cite{HeZ00L0YJ21}.  DPOS elects a limited number of trusted block producers to validate transactions and create blocks, resulting in high throughput and low latency. 
  EOSIO supports multiple programming languages, including C++, Python, and JavaScript, offering developers the flexibility to choose the language they are most comfortable with and that best suits their application requirements.
  This multi-language support enables developers to leverage their existing skills and expertise, facilitating smart contract creation and deployment. EOSIO smart contracts excel in handling high transaction throughput, with reported speeds of nearly 4,000 transactions per second (TPS). EOSIO emphasizes developer-friendliness, providing a web-based IDE that simplifies smart contract creation and deployment. This makes it suitable for DApps requiring fast, scalable transaction processing, such as social media or gaming applications. 
\end{itemize}

As the use of smart contracts becomes more widespread, ensuring their security has become paramount. In recent years, smart contracts have suffered from an increasing number of security issues, resulting in substantial financial losses and reputational damage. Consequently, it is crucial for both academics and industry professionals to focus on smart contract security and develop effective techniques and tools to safeguard these systems. 

\begin{figure}[t]
\centering
\includegraphics[width=3.8in]{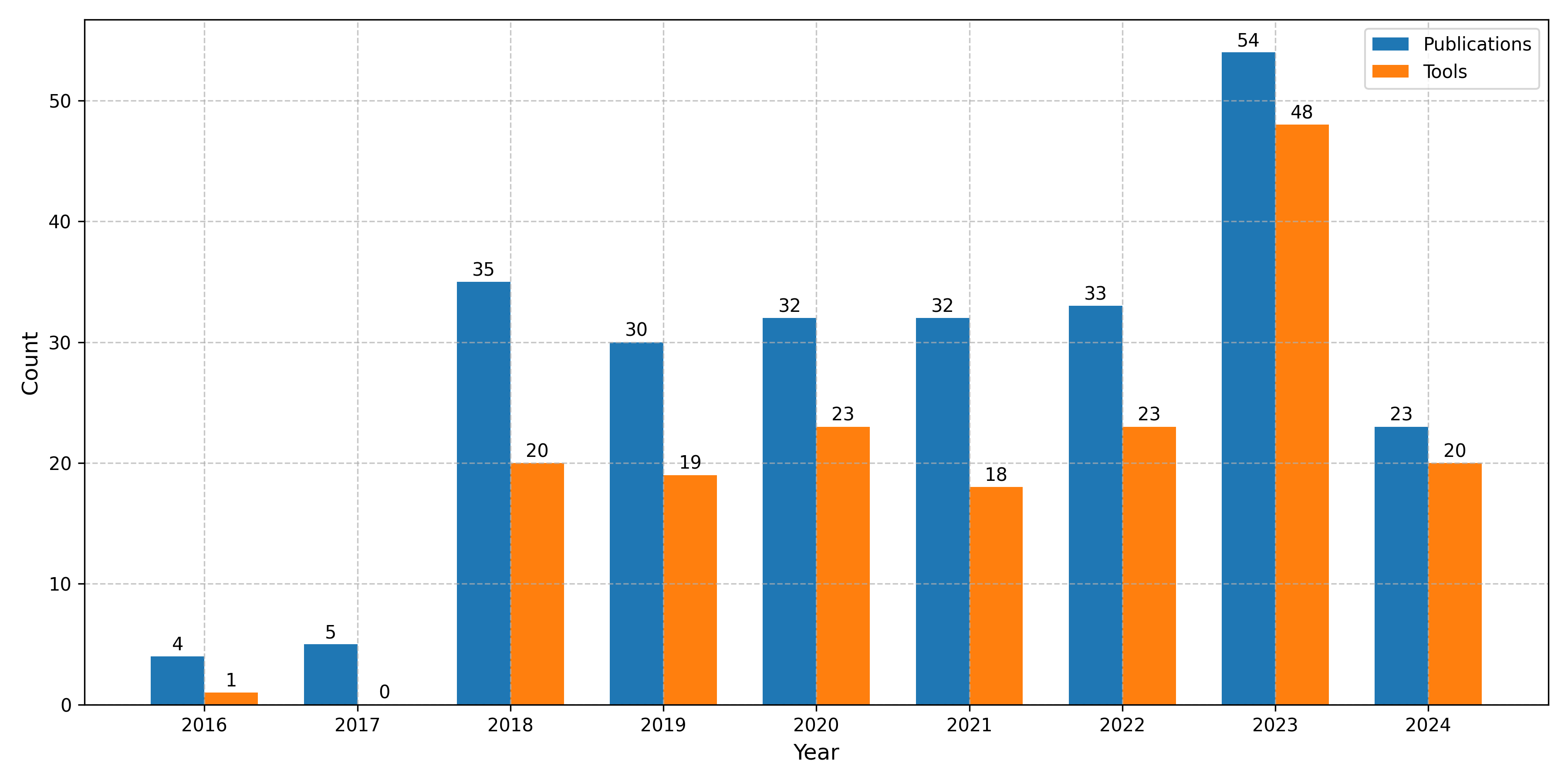}
\caption{Publications and Tools Over the Year}
\label{fig_statistics}
\end{figure}

Figure \ref{fig_statistics} illustrates the trend in smart contract security publications and tools from 2016 to July 2024. The data shows a marked increase in both publications and tool development from 2016 to 2018. Subsequently, the numbers stabilized between 2018 and 2022. A second notable increase occurred in 2023. 
This consistent growth in both academic publications and tools can be attributed to advancements in technologies like AI, the emergence of new vulnerabilities, and the rising value of smart contracts. These factors present new challenges and opportunities for security research. The growing importance of smart contracts in managing valuable and sensitive operations intensifies the need for robust security measures, further driving research and tool development in this field.

\subsection{Survey Methodology}
\label{survey_methodology}
The objective of our survey is to provide a comprehensive analysis of different approaches to dealing with vulnerable smart contracts. To achieve this, we have formulated a set of \textit{Research Questions (RQ)} that outline the scope of our review:

\begin{itemize}[leftmargin=*] 
  \item {RQ1: [Vulnerabilities]} What are the common vulnerabilities that exist in smart contracts, and how can we classify them based on their characteristics?
  \item {RQ2: [Attacks]} How do attackers exploit these vulnerabilities, and what are the consequences of these attacks on smart contracts?
  \item {RQ3: [Defenses]} What defense methodologies are available to protect smart contracts against attacks, and how do these defenses mitigate the risks associated with vulnerabilities?
  \item {RQ4: [Effectiveness]} How effective are the existing vulnerability analysis tools in identifying and detecting vulnerabilities in Solidity smart contracts, and what are the strengths and limitations of these tools in terms of accuracy, performance, and coverage?
  \item {RQ5: [Tools and Benchmark]} How do we select representative vulnerability analysis tools for evaluation, and how can we create a standardized benchmark of vulnerable and correct smart contract cases to evaluate the effectiveness of these tools?
\end{itemize}

By addressing these research questions, our survey aims to provide comprehensive insights into the vulnerabilities, attacks, defense mechanisms, effectiveness of analysis tools, and the creation of a benchmark for smart contracts. To achieve this, we employed a rigorous methodology to ensure comprehensive coverage and relevance of the selected literature. The process began with defining clear inclusion and exclusion criteria to guide our literature selection. These criteria were based on the relevance to smart contracts, publication date (papers published between 2015 and July 2024 were considered), and the credibility of the publication source.

Our search strategy involved querying major academic databases and conference proceedings, using keywords related to smart contracts. We searched through databases such as IEEE Xplore, ACM Digital Library, Google Scholar, Springer Link, and Scopus. We employed the following keywords and search terms: \textit{``blockchain" AND ``smart contracts" AND ``vulnerabilities" OR ``security" OR ``attacks"}. This initial search yielded a substantial number of results, as tabulated in Table \ref{tab:academic_survey}.


\begin{table}[ht]
    \centering
    \footnotesize
    \caption{Summary of Papers by Publication Platforms}
    \label{tab:academic_survey}
    \begin{tabular}{lr}
        \toprule
        \textbf{Publisher} & \textbf{Primary Results} \\
        \midrule
        Google Scholar & 97,900 \\
        IEEE Xplore Digital Library & 273,574 \\
        ACM Digital Library & 123,486 \\
        Springer Link & 936,311 \\
        Scopus & 5,004 \\
        \bottomrule
    \end{tabular}
\end{table}

The screening process generally consists of two stages: a title and abstract screening stage, followed by a full-text screening stage \cite{polanin2019best}. Once we have identified the relevant papers, we assess their quality based on predefined inclusion criteria. 
This generally involves assessing the relevance, methodology, validity, reliability, credibility, clarity, and impact. The selected papers should address the topic of smart contracts in a meaningful way and provide valuable insights for our investigation. 
The methodology employed should be suitable for addressing our research questions, transparent, and capable of being replicated. The results should be accurate and consistent, and the conclusions drawn should be well-supported by the data. The authors should present their ideas clearly and concisely, with the paper being free of errors and inconsistencies. Furthermore, the paper should have made a significant contribution to the field of smart contracts.

After evaluating these quality attributes, we have collected 308 related papers from 72 conference proceedings (including CCS, IEEE S\&P, NDSS, USENIX, Euro S\&P, Financial Crypto, ASIA CCS, ICSE, ASE) and 11 journals (including IEEE TSE, IEEE COMST), as well as relevant preprints. The next step involved extracting data from these papers, focusing on their research questions, methodologies, findings, and conclusions. This data extraction was conducted systematically to gather comprehensive insights from each study.
The extracted data and our analysis are documented in an online repository (\url{https://github.com/WeiZ-boot/survey-on-smart-contract-vulnerability}).

\section{Vulnerability in Smart Contracts}
\label{vulnerability}

\begin{figure}[!ht]
    \centering
    \includegraphics[width=5in]{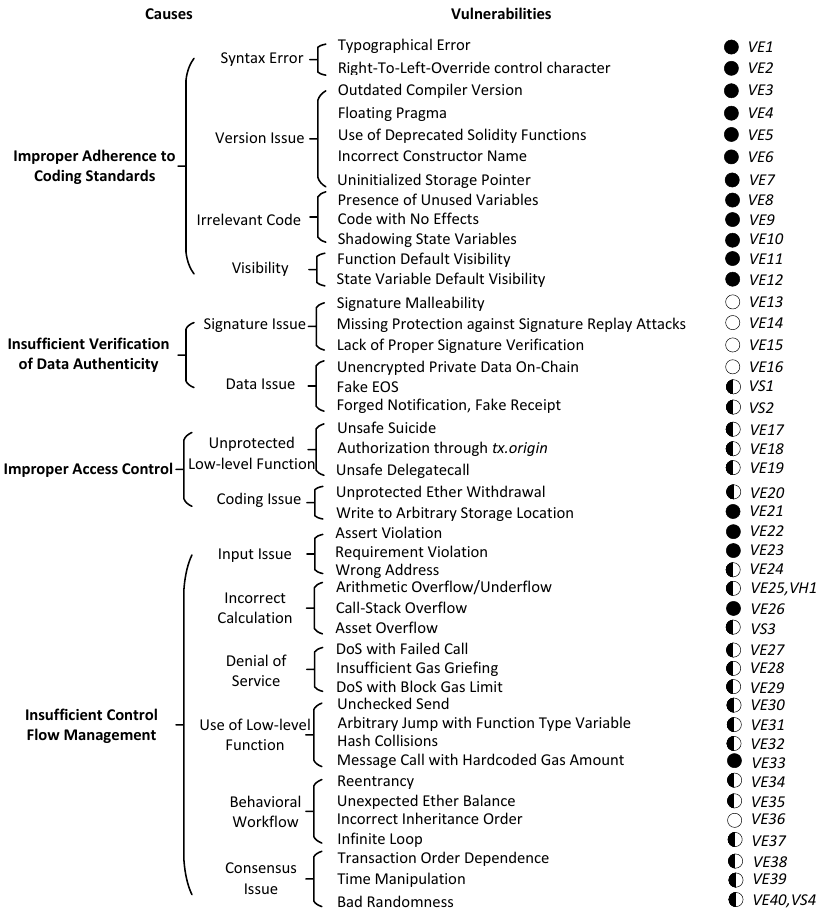}
    \caption{A classification of smart contracts vulnerabilities and their causes, where $\CIRCLE$  means the vulnerability has already been solved,  $\Circle$ means the vulnerability is widely discussed (not solved), and $\LEFTcircle$ means the vulnerability has been fixed by specific approaches or tools. \textit{VE} means vulnerabilities of Ethereum, \textit{VH} means vulnerabilities of HF, \textit{VS} means vulnerabilities of ESOIO.}
    \label{fig_classification}
\end{figure}

This section presents a robust methodology for effectively identifying and categorizing vulnerabilities in blockchain smart contracts. To ensure comprehensive coverage of various vulnerability types, our investigation is conducted from multiple perspectives. Specifically, for Ethereum smart contracts, we primarily reference two key projects: the Decentralized Application Security Project (DASP) \footnote{https://dasp.co/} and the Smart Contract Weakness Classification (SWC) Registry \footnote{https://swcregistry.io/}. DASP lists the top 10 smart contract vulnerabilities, while the SWC Registry details 37 vulnerabilities, each providing unique insights into specific vulnerability types.

Additionally, we also draw insights from several related research papers, including \cite{AtzeiBCLZ18, LvWWZ21, HeZ00L0YJ21, perez2021smart, hu2021comprehensive, rameder2022review}. By diligently collecting and examining all identified smart contract vulnerabilities and their respective causes, our aim is to establish a comprehensive methodology for categorizing the root causes of vulnerabilities in blockchain smart contracts. This methodology is built upon the well-established Common Weakness Enumeration (CWE) rules and effectively identifies four primary root causes of vulnerabilities: coding standards, data authenticity, access control, and control flow management. 

To illustrate our classification framework, Figure \ref{fig_classification} provides a visual representation of smart contract vulnerabilities based on their root causes and corresponding secondary causes. Our classification framework encompasses 14 distinct secondary causes, which are associated with 40 specific vulnerabilities found in Ethereum, Hyperledger Fabric (HF), and EOSIO. Furthermore, the figure also indicates the status of each vulnerability, indicating whether it has been eliminated, can be mitigated by specific methods, or remains unsolved. 
It is important to note that our study does not encompass Bitcoin smart contract security. This is because, unlike the aforementioned platforms, researchers have primarily focused on studying the construction and capabilities of Bitcoin's scripting language. Due to its inherent limitations, there is no widespread and unified smart contract language in Bitcoin, thus rendering the typical smart contract vulnerabilities less pertinent to its ecosystem. 

Recognizing the lack of consistency in the naming and definition of vulnerabilities across various studies, we have taken measures to ensure clarity and uniformity in our analysis. As part of our research, we have standardized the names of these vulnerabilities. Furthermore, we offer detailed explanations for each vulnerability to promote better comprehension and facilitate future research and analysis.

\subsection{Improper Adherence to Coding Standards}
This type of weakness happens when a smart contract is not developed in accordance with established coding rules and best practices. This issue often arises due to the relative novelty of programming languages used for smart contracts, leading to a shortage of experienced developers in the domain. Additionally, some developers may possess insufficient knowledge of language-specific coding standards, leading to errors and vulnerabilities in smart contract implementation. Improper adherence to coding standards can manifest in several ways, including:

\subsubsection{Syntax Errors (VE1, VE2)}
These errors occur when the code violates the syntax rules of the programming language, such as spelling and punctuation flaws. Two specific examples of syntax errors are \textit{typographical error (VE1)} and \textit{Right-To-Left-Override control character (VE2)}. \textit{VE1} refers to a typographical error in the code where an incorrect operator is used. \textit{VE2} involves the misuse of the U+202E unicode character, known as the Right-To-Left-Override control character. Both of these vulnerabilities can be mitigated by following best practices and employing preventive measures. For mitigating \textit{VE1}, pre-condition checks and proper code review processes can help identify typographical errors and incorrect operators, ensuring the code functions as intended. 
This method is supported by the work of Khajwal \cite{khajwal2023fast}, who demonstrates the effectiveness of pre-conditions in identifying typographical errors in programming. Additionally, thorough code review processes, as outlined by Sykes \cite{sykes2023seeing}, can serve as an effective measure to detect and rectify typographical errors. Regarding \textit{VE2}, using reliable libraries such as SafeMath can provide a robust foundation for performing secure mathematical operations, reducing the risk of errors and vulnerabilities.

\subsubsection{Version Issues (VE3, VE4, VE5, VE6, VE7)} 
Version issues in smart contracts can arise due to the rapid progress and updates in smart contract technology, including changes in compiler versions. When developers write code using outdated or deprecated functions, operators, or coding standards in a new compiler version, it can result in unexpected behaviors and potentially exploitable states. This category of vulnerabilities includes five specific flaws: \textit{outdated compiler version (VE3)}, \textit{Floating pragma (VE4)}, \textit{use of deprecated Solidity functions (VE5)}, \textit{incorrect constructor name (VE6)}, and \textit{uninitialized storage pointer (VE7)}. To avoid these vulnerabilities, it is essential to stay updated on the latest version of the compiler and adhere to the recommended coding standards and best practices. Using a recent version of the compiler ensures compatibility with the latest language features, bug fixes, and security enhancements. Additionally, developers should review and update their codebase to use the recommended functions, constructors, and storage initialization techniques specified in the updated compiler version.

\subsubsection{Irrelevant Code (VE8, VE9, VE10)}
Irrelevant code in smart contracts refers to code that is not essential for the execution or functionality of the contract. While this code may not directly impact the correctness of the contract, it can introduce security vulnerabilities or make them harder to detect. It is not uncommon for programming code to contain unused or shadowing parts. This category of vulnerabilities includes three specific flaws: \textit{presence of unused variables (VE8)}, \textit{code with no effects (VE9)}, and \textit{shadowing state variables (VE10)}. To mitigate these vulnerabilities, it is important for contract writers to thoroughly test the functionality and behavior of the code before deployment. This includes testing each intended behavior of the contract to ensure that it works as intended and does not contain irrelevant or unused code. By conducting comprehensive testing and code reviews, contract writers can reduce the risk of vulnerabilities introduced by irrelevant code in smart contracts.

\subsubsection{Visibility (VE11, VE12)}
Solidity provides access control labels for functions and variables in smart contracts, namely \textit{public}, \textit{external}, \textit{private}, or \textit{internal}. Each visibility label determines who can access or call specific functions or variables. The default visibility setting in Solidity is \textit{public}, which means that if the contract writer does not explicitly specify the visibility, functions and variables will be treated as \textit{public} by default. Forgetting to set the appropriate visibility for a function or variable can lead to two vulnerabilities: \textit{function default visibility (VE11)} and \textit{state variable default visibility (VE12)}. To mitigate these risks, contract writers should carefully consider the suitable visibility for each function and variable. Implementing pre-condition checks is crucial to ensuring that only authorized parties have access to critical functions or sensitive state variables.

\subsection{Insufficient Verification of Data Authenticity}
This type of weakness occurs when systems fail to properly verify the origin or authenticity of data, which can allow attackers to manipulate or access sensitive information. This can lead to a wide range of security issues, including \textit{cryptographic signatures} and \textit{cryptographic data}.

\subsubsection{Cryptographic Signatures (VE13, VE14, VE15)}:
Cryptographic signatures play a crucial role in validating the authenticity and integrity of data within blockchain systems. In Ethereum (and Bitcoin), the Elliptic Curve Digital Signature Algorithm (ECDSA) is commonly used for cryptographic signature generation and verification. However, there are certain vulnerabilities related to cryptographic signatures that can be exploited by attackers: \textit{signature malleability (VE13)}, \textit{missing protection against signature replay attacks (VE14)}, and \textit{lack of proper signature verification (VE15)}. To mitigate these vulnerabilities, it is of utmost importance to implement robust signature verification mechanisms. In the context of Ethereum, the built-in function \textit{ecrecover()} serves as a valuable tool for verifying ECDSA signatures. However, it is crucial to exercise diligence and precision when utilizing this function to ensure the thorough validation of signature integrity and authenticity. By doing so, we can effectively prevent issues such as signature malleability, replay attacks, and inadequate verification.

\subsubsection{Cryptographic Data (VE16, VS1, VS2)}
This type of vulnerability in smart contracts refers to situations where sensitive data, despite being marked as \textit{private}, can still be accessed by unauthorized parties. This vulnerability arises due to the inherent transparency of blockchain transactions, which allows the content of transactions to be readable by anyone.  Attackers can easily access and acquire the data stored in the contract, leading to significant financial losses for the contract creator and participants. The vulnerabilities associated with cryptographic data include \textit{unencrypted private data on-chain (VE16)}, \textit{fake EOS (VS1)}, and \textit{forged notification, fake receipt (VS2)}. To address these vulnerabilities, it is crucial to prioritize the proper encryption of sensitive data before storing it on-chain. This ensures that the data remains secure and protected. Furthermore, developers should exercise caution when dealing with contracts that involve private data, taking extra care to thoroughly test their contracts. By doing so, they can effectively minimize the risk of falling victim to honeypots or other malicious schemes.

\subsection{Improper Access Control}
This type of weakness arises when unauthorized users gain access to a contract and can perform actions that they should not be allowed to. Such vulnerabilities can have significant repercussions for the smart contract ecosystem, including financial losses and other adverse outcomes. Improper access control vulnerabilities manifest in two primary forms: \textit{unprotected low-level function} and \textit{coding issues}. Addressing these vulnerabilities is crucial to maintaining the security and integrity of smart contracts.

\subsubsection{Unprotected Low-level Function (VE17, VE18, VE19)}
Users can utilize Solidity's low-level functions \textit{SELFDESTRUCT}, \textit{tx.origin}, and \textit{DELEGATECALL} to control contracts. These low-level functions provide powerful capabilities but can be easily abused by malicious users if not used with caution. The following are the vulnerabilities associated with unprotected low-level functions:

\begin{itemize}[leftmargin=*]
  \item \textit{Unsafe suicide (VE17)} The \textit{SELFDESTRUCT} function allows a contract to be removed from the blockchain, returning any remaining Ether to a designated target address. While this can be useful in certain scenarios, it carries risks. If Ether is sent to a contract that has self-destructed, the funds will be permanently lost and cannot be recovered. Consequently, it is imperative to exercise caution when utilizing the \textit{SELFDESTRUCT} function. Developers should carefully consider the variables and conditions involved before employing this function. It is crucial to avoid referencing contract addresses or funds that could be manipulated in a way that leads to unintended loss of Ether. By adhering to these precautions, developers can mitigate the risk of irreversible Ether loss and ensure the integrity of their smart contracts.
  
  \item \textit{Authorization through tx.origin (VE18)} The \textit{tx.origin} variable represents the address that initiated a transaction, while \textit{msg.sender} represents the immediate invoker of a function. Relying on \textit{tx.origin} for authorization can lead to a vulnerability known as ``Authorization through tx.origin" (VE18). When one contract calls another contract, \textit{tx.origin} does not represent the calling address but rather the original initiator of the transaction. This can result in funds being transferred to the wrong address. To mitigate this vulnerability, it is recommended to use \textit{msg.sender} instead of \textit{tx.origin} for authorization checks.
  
  \item \textit{Unsafe delegatecall (VE19)} This vulnerability arises from the \textit{DELEGATECALL} instruction, which allows third-party code to be executed within the context of a current contract. This vulnerability, also known as ``Delegatecall to Untrusted Callee" (VE19), can be exploited by attackers to take control of another contract, especially in proxy contracts where code can be dynamically loaded from different addresses at runtime. If an attacker can manipulate the address used in \textit{DELEGATECALL}, they may modify storage or execute malicious code, leading to unauthorized actions such as fund theft or contract destruction.
\end{itemize}

\subsubsection{Coding Issues (VE20, VE21)}
Due to unintentionally exposing some functions, malicious parties can withdraw some or all Ether from the contract account. This type of flaw leads to \textit{unprotected Ether withdrawal (VE20)} and \textit{write to arbitrary storage location (VE21)}. In \textit{VE20}, contract writers wrongly name a function that is a constructor, and then the contract can be re-initialized by anyone. In \textit{VE21}, malicious users can write to sensitive storage locations, which can overwrite the original content and change the contracts. This type of flaw can be avoided by carefully designing the code or structure.

\subsection{Insufficient Control Flow Management}
This type of weakness occurs when attackers can exploit the openness of the public blockchain to gain control over the program's execution in unexpected ways. This weakness can manifest in various forms as follows:

\subsubsection{Improper Input (VE22, VE23, VE24)} Due to error handling in EVM, improper input can cause \textit{assert violation (VE22)}, \textit{requirement violation (VE23)} and \textit{wrong address (VE24)}. The Solidity assert(), require(), as guard functions are introduced to improve the readability of contract code. However, assert() and require() require strong logical conditions, and improper use will cause errors. Moreover, the length of a contract address should be 40 hexadecimal characters. If the address length is incorrect, the contract can still be deployed without any warning from the compiler. Ethereum automatically registers a new address that is owned by nobody, and any Ether sent to this address becomes inaccessible. To mitigate these vulnerabilities, it is important to carefully handle input validation, use assert() and require() with appropriate conditions, and validate the format and length of contract addresses. 

\subsubsection{Incorrect Calculation (VE25, VH1, VE26, VS3)} This type of weakness happens when contracts perform a calculation that generates incorrect results and that may lead to a larger security issue such as arbitrary code execution. \textit{Arithmetic overflow/underflow (VE25, VH1)} is the most common error in software, and it both happens in Ethereum and HF. \textit{Call-stack overflow (VE26)} occurs due to the Ethereum Virtual Machine (EVM) imposing a limit on the depth of the call-stack, allowing a maximum of 1024 nested function calls. If an attacker successfully reaches this limit by repeatedly invoking functions, it can result in a call-stack overflow vulnerability. Once the call-stack reaches its maximum depth, subsequent instructions, such as the \textit{send} instruction, will fail. \textit{Asset overflow (VS3)} specifically pertains to the EOSIO blockchain. It occurs when there is an overflow in the asset type, which represents token balances and other asset values on the EOSIO platform.

\begin{itemize}[leftmargin=*]
 \item \textit{Arithmetic overflow/underflow (VE25)} This vulnerability, commonly known in software programming, is not specific to smart contracts. It occurs when an arithmetic operation produces a value that exceeds the maximum or minimum range of integer representation. In Ethereum contracts, this vulnerability arises due to the behavior of the EVM's integer arithmetic and the lack of automatic checks for arithmetical correctness. For instance, if the result of an addition operation surpasses the maximum value representable by a specific integer type, it wraps around to a lower-than-expected value without raising an error or warning. Research by Torres et al. \cite{TorresSS18} has identified over 42,000 contracts, particularly ERC-20 Token contracts, vulnerable to arithmetic overflow/underflow. To mitigate this issue, it is advisable to employ libraries such as SafeMath. These libraries offer secure implementations of arithmetic operations with built-in checks to prevent overflow and underflow errors.
 
\end{itemize}

\subsubsection{Denial of Service (VE27, VE28, VE29)} Denial of Service vulnerabilities can affect smart contracts and result in exceptions that may lead to undesirable consequences such as contract lock-ups or freezing of funds. There are multiple ways in which DoS attacks can be carried out, with two primary incentives: failed calls and gas consumption. One type of DoS vulnerability is \textit{DoS with failed call (VE27)}, where an external call, whether accidental or deliberate, fails. This vulnerability is particularly relevant in payment scenarios where multiple calls are executed within a single transaction. The failure of an external call can disrupt the intended flow of the contract, potentially leading to undesired consequences. Another DoS problem can be concluded as gas-related vulnerabilities. Gas is a mechanism designed to prevent resource abuse, and each operation within a smart contract consumes a certain amount of gas. One such vulnerability is \textit{insufficient gas griefing (VE28)}, which occurs when there is not enough gas to support an external call, causing the transaction to be reverted. Another gas-related vulnerability is \textit{DoS with block gas limit (VE29)}, also known as \textit{DoS with Unbounded Operations}. Mitigating these DoS vulnerabilities requires careful design and consideration of gas usage.

\begin{itemize}[leftmargin=*]
  \item \textit{DoS with block gas limit (VE29)}. To prevent resource abuse, especially DoS attacks, Ethereum proposes gas mechanism to charge an execution fee from each operation paid by transaction senders. To further safeguard the network, each block has a predetermined maximum amount of gas that can be consumed, known as the \textit{Block Gas Limit}. The gas consumption of a transaction must be less than or equal to the Block Gas Limit; otherwise, the transaction will fail to execute and any changes made during its execution will be rolled back. This ensures that no single transaction monopolizes excessive resources within a block, promoting fair usage and preventing DoS attacks that could overwhelm the network. It is important to note that different EVM instructions have varying gas costs. Some operations, such as ADD, AND, and POP, have relatively low gas costs, while others, like SSTORE, incur higher gas costs. This differentiation encourages efficient and responsible use of gas resources. 
\end{itemize}

\subsubsection{Use of Low-level Function (VE30, VE31, VE32, VE33)} 
Solidity's low-level functions, such as \textit{call}, \textit{transfer}, \textit{send}, \textit{mstore} and \textit{abi.encodePacked}, provide users with control and flexibility when interacting with smart contracts.  However, improper use of these low-level functions can introduce unexpected behavior and vulnerabilities into the contract's program logic. \textit{Unchecked send (VE30)} arises when developers make the common assumption that a particular function will never fail. If a malicious user can force the function to fail, it can leave the contract in an unexpected state. The use of \textit{transfer} and \textit{send} functions can lead to \textit{message call with hardcoded gas amount (VE33)} when the gas cost significantly changes, such as during hard forks or network upgrades. The \textit{mstore} function can introduce the \textit{arbitrary jump with function type variable (VE31)} vulnerability if an attacker can manipulate a function type variable to point to any code instruction. This can potentially allow an attacker to execute arbitrary code and compromise the integrity of the contract. \textit{Hash collisions (VE32)} occur when the \textit{abi.encodePacked} function is used incorrectly with multiple variable length arguments. To mitigate these vulnerabilities, it is crucial to carefully review and validate the usage of low-level functions, handle exceptions appropriately, account for potential changes in gas costs, and implement robust input validation and verification mechanisms.

\begin{itemize}[leftmargin=*]
  \item \textit{Unchecked send (VE30)} This vulnerability is also described as \textit{unhandled exceptions}, \textit{exception disorder}, or \textit{unchecked low-level call}. This issue happens when the call fails accidentally or an attacker forces the call to fail.
  In some cases, developers may include code to check the success of the call, but they neglect to handle the exceptions properly. As a result, funds intended for transfer may not reach the intended recipient. This vulnerability stems from the inconsistent exception-handling behavior in Solidity, which can lead to unexpected outcomes if not handled correctly. 
\end{itemize}

\subsubsection{Improper Behavioral Workflow (VE34, VE35, VE36, VE37)} 
It refers to vulnerabilities that arise when the expected order or sequence of operations within a smart contract is manipulated by malicious users, leading to unexpected states or undesired behavior. These vulnerabilities include \textit{reentrancy (VE34)}, \textit{unexpected Ether balance (VE35)}, \textit{incorrect inheritance order (VE36)}, and \textit{infinite loop (VE37)}.
In \textit{reentrancy (VE34)}, a malicious contract calls back into the calling contract before the first invocation of the function is over. \textit{Unexpected Ether balance (VE35)} occurs when malicious users intentionally send funds to a contract in a specific manner to disrupt its intended behavior or cause a denial-of-service condition. By manipulating the contract's ether balance, attackers can affect the contract's functionality and, in extreme cases, render it unusable. \textit{Incorrect inheritance order (VE36)} is a vulnerability that arises from the improper ordering of contract inheritance. In Solidity, contracts can inherit from other contracts, and the order of inheritance can impact the behavior and functionality of the derived contract. Malicious users can manipulate the inheritance order to achieve unexpected outcomes and potentially exploit vulnerabilities. \textit{Infinite loop (VE37)} refers to a vulnerability where a contract falls into an infinite loop, leading to non-termination of contract execution. This vulnerability is often associated with the fallback function being incorrectly invoked, resulting in repeated and endless execution of the fallback code. This can exhaust the contract's gas and disrupt the intended functionality of the contract.

\begin{figure}[!h]
    \centering
    \includegraphics[width=2.5in]{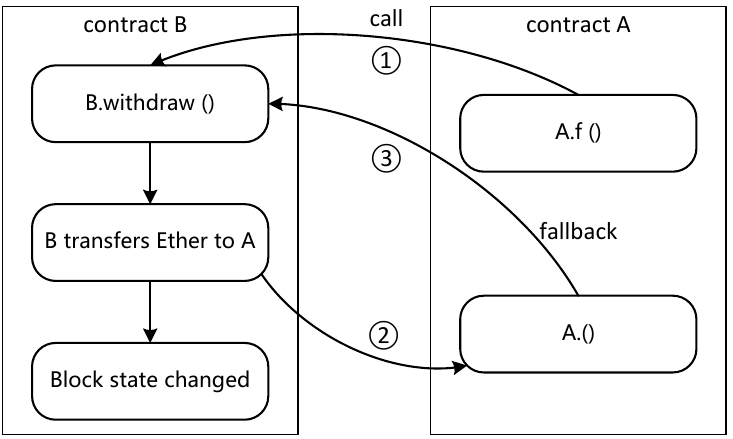}
    \caption{Reentrancy attack flow}
    \label{reentrancy-vul}
  \end{figure}

\begin{itemize}[leftmargin=*]
  \item \textit{Reentrancy (VE34)} This vulnerability occurs when a contract invokes a function from an external contract, and the called contract has sufficient gas to invoke a callback into the calling contract. This creates a loop where the called contract re-enters the calling contract before the initial invocation is completed. Malicious attackers can exploit this vulnerability to manipulate the execution flow and potentially exploit vulnerabilities present in the contract. It is crucial to carefully review and secure contract interactions to prevent re-entrance attacks and ensure the integrity and security of the smart contract system. To illustrate how this vulnerability is exploited, we provide an example shown in Figure \ref{reentrancy-vul}. The example involves two contracts: Contract A, representing the attacker, and Contract B, representing the victim. The vulnerability is demonstrated in three steps:
  \textcircled{1} Contract A invokes the function f to call the function to withdraw Contract B; \textcircled{2} where B subsequently transfers some ether to A (msg.sender represents the address of contract A). This step follows a common withdrawal pattern \cite{Withdrawal}; \textcircled{3} As part of the ether transfer, Contract B invokes the fallback function of Contract A, allowing Contract A to specify any function (``call to the unknown"). In this scenario, Contract A re-enters Contract B through the fallback function, which has no specific name. This re-entrance allows Contract A to repeatedly call the fallback function and potentially exploit vulnerabilities within Contract B.
\end{itemize}

  
\subsubsection{Consensus Issues (VE38, VE39, VE40, VS4)} In the blockchain, the synchronization of subsequent blocks with the majority of the network relies on following a consensus protocol, such as Proof of Work (PoW) or Proof of Stake (PoS). This consensus protocol allows network participants sufficient time to reach an agreement on which transactions should be included in the blocks. However, the synchronization process itself introduces vulnerabilities that can be exploited by attackers. These vulnerabilities include \textit{transaction order dependency (VE38)} (TOD), \textit{time manipulation (VE39)}, and \textit{bad randomness (VE40, VS4)}.

\begin{itemize}[leftmargin=*]
  \item \textit{TOD (VE38)}: This vulnerability, also known as Frontrunning, occurs due to the prioritization mechanism for transactions in blockchain blocks. Miners have the ability to choose which transactions to include in a block and the order in which they are arranged. Since transactions are often prioritized based on gas price, a malicious miner who can see and react to transactions before they are mined may manipulate the transaction order to their advantage. By varying the order of transactions and manipulating the output of the contract, they can manage undesirable outcomes or financial losses for users. 
  \item \textit{Time manipulation (VE39)}:  This vulnerability arises when smart contracts rely on the timestamp information from blocks to perform certain functions. In Solidity, the current timestamp can be obtained using \textit{block.stamp} or \textit{now}. However, this timestamp value can be manipulated by miners. If a contract's functionality is dependent on the timestamp, miners can profit by choosing a suitable timestamp to manipulate the contract's behavior. This vulnerability is also referred to as ``block values as a proxy for time".
  \item \textit{Bad randomness (VE40)}: This vulnerability refers to vulnerabilities in the generation of random numbers within smart contracts. Random numbers are often used to make decisions or determine outcomes. If the random number generation process is flawed, malicious actors may be able to predict the outcome of the contract and exploit it. One example of bad randomness is the use of a predictable seed value for the random number generator. If an attacker can guess or determine the seed value, they can predict the generated random numbers and manipulate the contract accordingly to their advantage.
\end{itemize}

\subsection{Common Vulnerability Ranking}
\begin{figure}[!ht]
    \centering
    \includegraphics[width=4in]{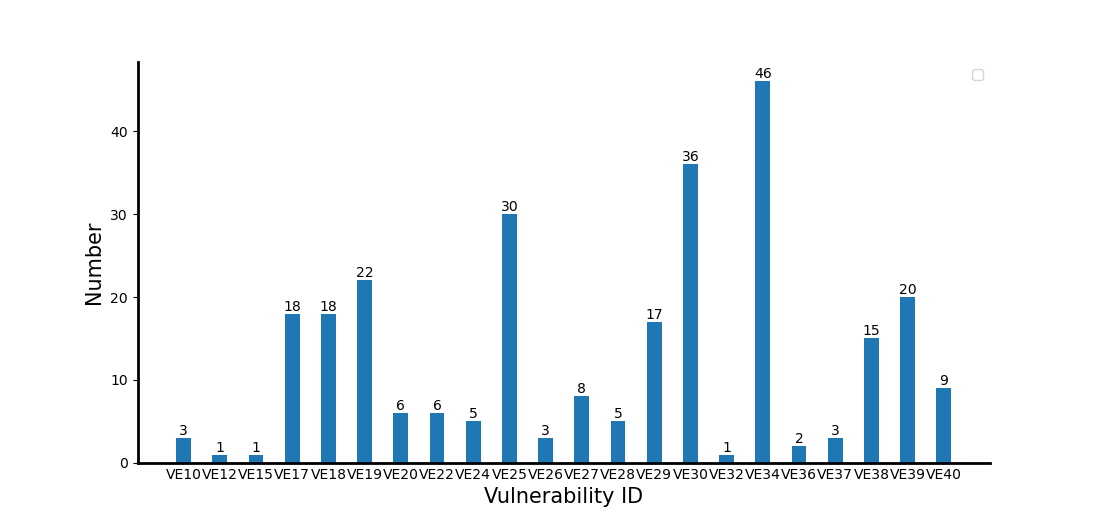}
    \caption{Vulnerabilities Frequency Statistics}
    \label{vulner_num}
\end{figure}

Since the last update of the DASP (Decentralized Application Security Project) in 2018, we have created a new list of the top 10 smart contract vulnerabilities that pose significant risks to the security and functionality of contracts. This list is based on the frequency of occurrence among various analysis tools available\footnote{\url{https://github.com/WeiZ-boot/survey-on-smart-contract-vulnerability}}. Figure \ref{vulner_num} provides statistics on the occurrence of 22 vulnerability categories. The top 10 smart contract vulnerabilities are as follows:
    \textit{reentrancy (VE34)}, 
    \textit{arithmetic overflow/underflow (VE25)}, 
    \textit{DoS with block gas limit (VE29)}, 
    \textit{unsafe suicidal (VE17)}, 
    \textit{unsafe delegatecall (VE19)}, 
    \textit{unchecked send (VE30)}, 
    \textit{TOD (VE38)}, 
    \textit{time manipulation (VE39)}, 
    \textit{authorization through tx.origin (VE18)}, 
    and various other vulnerabilities. 
These vulnerabilities have been identified as the most common and high-risk issues that developers should prioritize when assessing the security of their smart contracts.

The \textit{reentrancy} vulnerability attracts significant attention from researchers due to its difficulty in detection and mitigation. The complex and decentralized nature of smart contracts makes it challenging to ensure the atomic execution of functions and proper handling of reentrant calls. 
\textit{Arithmetic overflow/underflow} is a common issue in software programs, particularly those written in low-level languages. 
Gas usage plays a crucial role in the EVM and the smart contract ecosystem, and accurately estimating the required gas for an operation can be difficult. Given that smart contract applications often handle sensitive data and substantial amounts of value, \textit{DoS with block gas limit} can lead to denial-of-service attacks, data corruption, and financial losses for users. 
\textit{Unsafe suicidal}, \textit{unsafe delegatecall}, and \textit{authorization through tx.origin} vulnerabilities occur when the access control of a smart contract is flawed. Access control determines which entities can interact with the contract and what actions they can perform. These vulnerabilities can result in security risks and potential financial losses when access control is compromised. 
\textit{Unchecked send} occurs when a contract fails to handle exceptions from failed calls appropriately. This vulnerability causes the smart contract to behave unexpectedly and compromises its secure operation. 
\textit{TOD}, \textit{time manipulation}, and \textit{bad randomness} vulnerabilities are related to consensus issues influenced by the blockchain network. Smart contracts are executed on the blockchain and must follow the same transaction order as the underlying blockchain. This means that even if a smart contract is designed to be resistant to \textit{TOD} and \textit{time manipulation}, it can still be vulnerable if the underlying blockchain is not resistant to these issues.

In this section, we performed a comprehensive analysis of the root causes of vulnerabilities in the smart contract domain and introduced a novel classification system to effectively categorize them. Furthermore, we conducted a statistical ranking of the most frequently encountered vulnerabilities based on existing research. These findings offer conclusive and precise answers that effectively address our research question, \textbf{RQ1}, as outlined in Section \ref{survey_methodology}. 

By gaining a good understanding of these vulnerabilities in smart contracts and their ranking, developers can effectively allocate their time and resources, prioritizing the resolution of the most critical security concerns. Additionally, to fully comprehend the potential damage caused by these vulnerabilities, it is crucial to explore the common types of attacks that can exploit them. By carefully examining the relationship between vulnerabilities and attacks, developers can identify potential attack vectors and proactively implement robust measures to mitigate these risks.

\section{Attacks on Smart Contracts}
\label{attacks}
A recent survey \cite{zhou2023sok} reveals a rapid increase in the number of Solidity contracts over the past five years. This growth reflects the expanding range of smart contract applications across sectors such as DeFi \cite{berg2022empirical}, insurance, and lending platforms. 
Unfortunately, this growth has also led to an increase in the number of attackers exploiting vulnerabilities in smart contracts. Consequently, several high-profile attacks have occurred, resulting in substantial financial losses. 

\begin{figure}[!ht]
  \centering
  \includegraphics[width=5in]{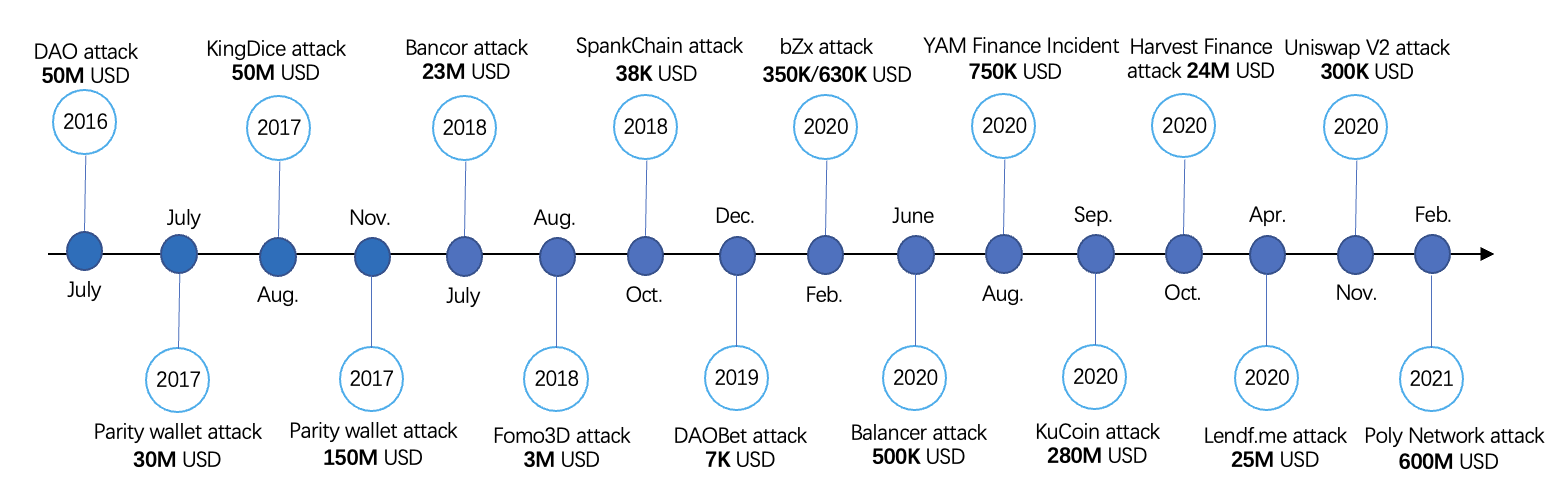}
  \caption{Several high-profile attacks from 2016 to 2021}
  \label{fig_attack_timeline}
\end{figure}

\begin{figure}[!ht]
    \centering
    \includegraphics[width=4.5in]{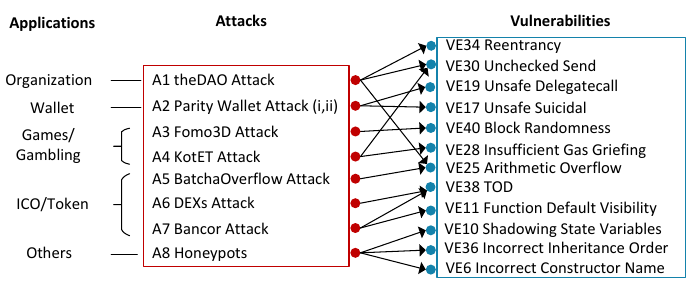}
    \caption{The relationships between attacks and vulnerabilities}
    \label{fig_attack}
\end{figure}

The most notable among these is the DAO attack in 2016, which led to millions of dollars in Ether being drained from the organization. Since then, smart contracts have experienced several high-profile attacks from 2016 to 2021, as shown in Figure \ref{fig_attack_timeline}. 
To gain a better understanding of these attacks, we have analyzed and identified eight representative attack patterns, as shown in Figure \ref{fig_attack}. This figure highlights the major application domains of smart contracts and the corresponding targeted attack patterns. Additionally, we conduct an examination of the vulnerabilities that contribute to these attacks. Through the classification of vulnerabilities based on known attack patterns, we can identify common weakness that require attention.
Following this analysis, we provide a comprehensive breakdown of each attack, shedding light on the specific vulnerabilities that contribute to these attacks.

\subsection{The DAO Attack (A1)}
The DAO (Decentralized Autonomous Organization) was a groundbreaking project launched on the Ethereum blockchain in 2016. It aimed to create a decentralized venture capital fund where participants could invest in projects and vote on funding decisions through the use of smart contracts.  The DAO project gained significant attention and raised over \$150 million USD in funding from investors. However, the DAO's success was short-lived as it fell victim to a critical security vulnerability. In June 2016, an attacker exploited a flaw in the DAO's smart contract code, allowing them to drain approximately \$60 million USD worth of ether from the organization \cite{AtzeiBC16}. 
This event, known as the DAO attack, was a significant setback for the Ethereum community and led to a contentious debate about the immutability of blockchain transactions and the need for a hard fork to recover the stolen funds. Ultimately, a hard fork was implemented to create a new version of the Ethereum blockchain that reversed the effects of the hack and returned the stolen funds to their rightful owners. Following the occurrence of The DAO hack, other smart contracts, such as Spankchain \cite{spankchain} and Lendf.me \cite{siliconangle}, also experienced losses due to comparable security vulnerabilities. Atzei et al. \cite{AtzeiBC17} provided two attack contract examples, called Mallory and Mallory2. In Mallory, attacker manipulates the control flow and exploits \textit{reentrancy (VE34)}. In Mallory2, attacker efficiently exploits \textit{arithmetic overflow/underflow (VE25)} and \textit{unchecked send (VE30)} vulnerabilities by using only two calls.

\subsection{Parity Wallet Attack (A2)}
The Parity Wallet attack encompasses two distinct incidents that occurred in 2017. During these incidents, the Parity Multisig Wallet suffered breaches, leading to substantial financial losses. The Parity Wallet is comprised of two components: a library contract and wallet contracts. The library contract contains the essential functions of a wallet, while the wallet contracts act as proxies that delegate calls to the library contract through the delegatecall mechanism. In both incidents, the vulnerabilities enabled the attackers to obtain unauthorized control over the wallets. The first incident is caused by exploiting \textit{unsafe delegatecal (VE19)} and more than 30M USD worth of ether is drained \cite{Palladino2017}. In this incident, the attacker initiated two transactions to manipulate each contract involved. The first transaction was aimed at gaining ownership of the victim's wallet contract.
The second incident on the Parity Multisig Wallet exploited the vulnerability known as \textit{suicide (VE17)} and resulted in the locking of more than \$280 million USD worth of funds \cite{Sergey}. In this incident, the attacker leveraged the delegatecall mechanism to initialize themselves as the owner of the wallet contract, similar to the first incident.

\subsection{Fomo3D Attacks (A3)}
The Fomo3D contract was an Ethereum game where participants could purchase keys using Ether and refer others to the game to earn more Ether. The objective of the game was to be the last participant to purchase a key before the timer expired, thereby winning the entire pot of Ether. The attacker purchased a ticket and then sent multiple transactions with high gas prices in rapid succession, effectively consuming a significant portion of the block's gas limit. This action caused other transactions related to Fomo3D, including the key purchases made by other participants, to be delayed or stuck in a pending state until the timer ran out.
The attackers gain an advantage by exploiting two vulnerabilities: \textit{bad randomness (VE40)} and \textit{DoS with block gas limit (V29)}.

\subsection{KotET Attack (A4)}
The King of the Ether Throne (KotET) contract was also a game contract where participants competed to win the throne and receive all the Ether held in the contract. This game contract was implemented as a contract account on the Ethereum blockchain.
The KotET attack, which occurred in February 2016 \cite{KotET}, exploited two vulnerabilities: \textit{unchecked send (VE30)} and \textit{insufficient gas griefing (VE28)}. In the attack, when the KotET contract attempted to transfer funds to another wallet contract, both contracts required sufficient gas to successfully process the transaction. However, if the wallet contract had insufficient gas, it would fail to complete the payment, resulting in the funds being returned to the KotET contract. Importantly, the KotET contract was not aware of the payment failure, and the latest player would be crowned as the King, while the compensation payment intended for the previous player would not be sent.

\subsection{BatchOverflow Attack (A5)}
The BatchOverflow attack, which occurred in April 2018, targeted the Beauty Ecosystem Coin (BEC) token. The attack exploited \textit{arithmetic overflow/underflow (VE25)} vulnerability to achieve an unauthorized increase in digital assets. This vulnerability resulted in the theft of BEC tokens and a temporary shutdown of the exchange platform. According to the blockchain security firm, PeckShield\footnote{https://peckshield.com/}, the BatchOverflow attack was not limited to the BEC token. According to the blockchain security firm PeckShield, they discovered similar integer overflow vulnerabilities in around 12 other token smart contracts. Some examples include SMT (proxyOverflow), UET (transferFlow), SCA (multiOverflow), HXG (burnOverflow), and others.

\subsection{Frontrunning Attack (A6,A7)}
\begin{figure}[htp]
	\centering
	\includegraphics[width=5in]{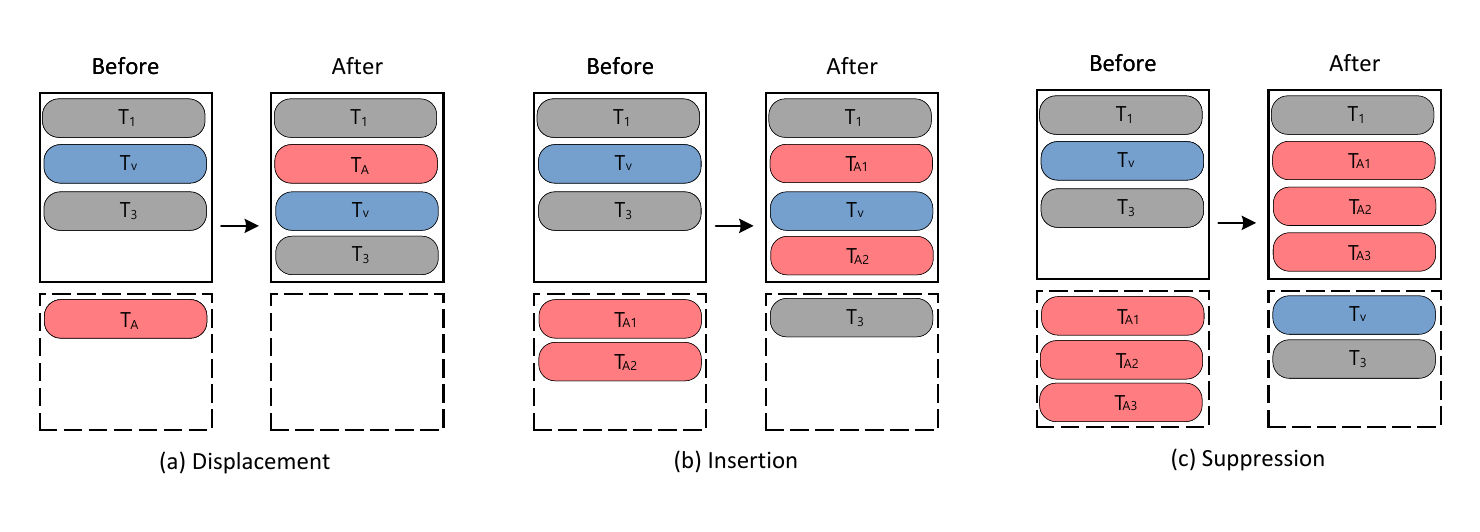}
	\caption{The three frontrunning attack cases}
	\label{fig_frAttack}
\end{figure}

Both the DEXs attack and the Bancor attack can be categorized as Frontrunning attacks. The concept of Frontrunning is not exclusive to blockchain and is commonly observed in traditional financial markets. Frontrunning involves manipulating financial markets by gaining undisclosed information about transactions beforehand \cite{TorresCS21}. This practice is generally deemed illegal in most countries. 
In the context of blockchain, every transaction is publicly visible in the pending pool before being included in a block. Miners, who have the authority to choose which transactions to include, often prioritize those with higher gas prices. This creates an opportunity for attackers to manipulate the transaction order and maximize their own profits. 
Frontrunning attacks exploit the vulnerability known as \textit{TOD (VE38)}, which is also referred to as the Frontrunning vulnerability. Eskandari et al. \cite{EskandariMC19} categorize this vulnerability into three types of attacks: displacement, insertion, and suppression attacks, as depicted in Figure \ref{fig_frAttack}.

\begin{itemize}[leftmargin=*]
  \item{\textbf{Displacement Attack}}: As highlighted in Figure \ref{fig_frAttack}(a), the attacker observes a profitable transaction $T_V$ from the victim and creates a new transaction $T_A$ with a higher gas price. The attacker then broadcasts $T_A$ to the network. When the miner assembles the transactions into a block, they prioritize including $T_A$ before $T_V$, effectively displacing the victim's transaction. As a result, the attacker can front-run the victim and potentially benefit from favorable market conditions.
  \item{\textbf{Insertion Attack (Sandwich Attack)}}: As highlighted in Figure \ref{fig_frAttack}(b), the attacker observes a profitable transaction $T_V$ and creates two transactions, $T_{A1}$ and $T_{A2}$. $T_{A1}$ has a higher gas price than $T_V$, and $T_{A2}$ has a lower gas price. The attacker broadcasts $T_{A1}$ and $T_{A2}$ to the network. When the miner assembles the transactions into a block, they place $T_A$ between $T_{A1}$ and $T_{A2}$, effectively sandwiching the victim's transaction. This allows the attacker to exploit price discrepancies or other favorable conditions.
  \item{\textbf{Suppression Attack}}:  As highlighted in Figure \ref{fig_frAttack}(c), the attacker observes a profitable transaction $T_V$ and decides to exclude it from the block. The attacker creates multiple transactions with higher gas prices than $T_V$ to fill up the block. When the miner assembles the transactions into a block, they intentionally leave out $T_V$, suppressing its execution. This prevents the victim from benefiting from the transaction and allows the attacker to potentially take advantage of market conditions. In this attack, \textit{DoS with block gas limit (VE29)} vulnerability is also present, as the attacker fills the block with high gas price transactions to consume the block gas limit.
\end{itemize}

\subsubsection{DEXs Attack (A6)} 
The decentralized exchange (DEX) is an exchange platform built on smart contracts where users can exchange their ERC-20 tokens for ether or other tokens. Frontrunning attacks have become a significant concern in DEXs. These attacks exploit the transaction order dependence vulnerability to manipulate the execution of transactions and gain an unfair advantage in trading. When a searcher detects a transaction that can yield a profit, it attempts to front-run that transaction by submitting its own transaction with higher gas fees to ensure its execution before the victim's transaction. The attacker's transaction is designed to take advantage of the anticipated price movement caused by the victim's transaction. 

For example, in the case of PancakeSwap DEX, the attacker's bot may observe a buyer trying to purchase 1 Corgicoin token. The bot quickly submits its own transaction to buy the Corgicoin token and then immediately sells it after someone else's transaction to maximize profits. Since these transactions occur within the same block, the attacker effectively front-runs the victim's transaction and benefits from the price movement caused by the victim's trade. These front-running attacks are performed by highly competitive bots that aim to maximize their profits by extracting Miner's Extractable Value (MEV). According to Flashbots \cite{flashbots}, searchers extracted approximately 691 million USD worth of value from Ethereum in January 2023 alone.

\subsubsection{Bancor Event (A7)}
The Bancor ICO is a decentralized exchange platform that allows users to create and trade their own tokens. During the audit of its exchange smart contract, two vulnerabilities were discovered: \textit{TOD (VE38)} and \textit{function default visibility (VE11)}. The attacker exploited the \textit{TOD} vulnerability, which enabled them to execute transactions ahead of others, resulting in a profit of 135,229 USD worth of ether \cite{Bancor}. Fortunately, no real-world attack occurred, but the vulnerability itself was identified and addressed.

\subsection{Honeypots (A8)}
Honeypots in the context of smart contracts are a type of fraudulent scheme that capitalizes on security vulnerabilities while employing deceptive tactics. The fundamental concept of a honeypot involves intentionally creating a smart contract that appears to possess an obvious flaw or vulnerability, enticing potential victims to exploit it for financial gain. However, the contract is deliberately designed in a way that when someone attempts to take advantage of the apparent vulnerability, they end up losing their funds instead. Honeypots manipulate human greed and the desire for quick profits, luring unsuspecting users into falling for the trap. The aim is to present an enticing opportunity for financial gain, while in reality, it is a trap orchestrated by malicious actors. 
According to Torres et al. \cite{TorresSS19}, honeypots typically take advantage of specific vulnerabilities to carry out their deceptive schemes such as \textit{incorrect constructor name (VE6)}, \textit{shadowing state variables (VE10)}, and \textit{incorrect inheritance order (VE36)}.

In this section, we have systematically examined eight common attack patterns and their associated lists of vulnerabilities. This analysis has allowed us to address the second research question, \textbf{RQ2}, as outlined in Section \ref{survey_methodology}, which focuses on how these common attacks exploit vulnerabilities and the consequences of such attacks. 
Our study of past attacks provides valuable insights into common vulnerabilities and attack patterns, as well as emerging trends in smart contract security. These include:

\begin{itemize}[leftmargin=*]
  \item \textbf{Increasing Complexity of Smart Contract Attacks.} The sophistication of smart contracts, with advanced features and integrations, has led attackers to develop more complex techniques for exploitation. These include attacks that leverage multiple vulnerabilities simultaneously or that exploit the interactions between different smart contracts.
  \item \textbf{Rise in Attacks Targeting DeFi Applications.} DeFi platforms, due to their high liquidity and rapid growth, have become lucrative targets for attackers. Vulnerabilities in these platforms can lead to significant financial losses, as seen in various high-profile incidents in recent years.
  \item \textbf{Potential Advent of AI-Enabled Attackers.} AI-enabled attacks could potentially identify vulnerabilities in smart contracts more quickly and exploit them before they are detected and remedied by developers or security teams. This includes the ability of AI systems to analyze vast amounts of code at an unprecedented speed, identify patterns and weaknesses overlooked by human analysis, and even adapt to changing security environments.
\end{itemize}

These insights are crucial for smart contract developers in enhancing the security of their contracts. While it may be challenging to completely eliminate all attacks on vulnerable smart contracts, there are steps that can be taken to minimize risks and improve contract security. By implementing best practices and employing appropriate security measures, developers can reduce the likelihood of their contracts being exploited by attackers. Building upon our understanding of attack patterns, we can now explore various defense methodologies in the next section.

\begin{figure}[t]
	\centering
	\includegraphics[width=5in]{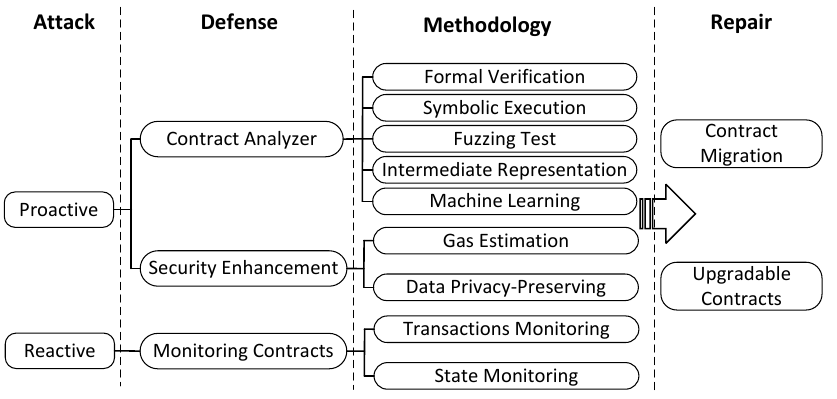}
	\caption{Research ideas for the defense of smart contracts}
	\label{fig_defense}
\end{figure}

\section{Defense Methodologies}
\label{defense}
Defense technologies evolve alongside the advancement of attacks, and there is a substantial body of work on security measures for smart contracts \cite{zhou2020ever, ChenPNX20, TorresCS21}. Figure \ref{fig_defense} provides an overview of a comprehensive set of research solutions for smart contract defense. Generally speaking, defense strategies for smart contracts can be categorized into two main groups: proactive and reactive. Proactive defense strategies involve taking preventive measures to mitigate attacks before they occur, while reactive defense strategies focus on responding to attacks after they have happened. Proactive defense strategies aim to address known attacks that have been previously identified and studied, with defense mechanisms developed specifically for them. These attacks can be prevented or mitigated using contract analyzers or security enhancements. Reactive defense strategies, on the other hand, pertain to attacks that have not been previously identified or studied, and for which defense mechanisms may not exist. Monitoring contracts play a crucial role in addressing such attacks. However, it is important to note that identifying vulnerabilities and deploying monitoring contracts alone are insufficient for effective smart contract defenses. The repair of vulnerable smart contracts is also a key technology in the defense process \cite{RodlerLKD21, TorresJS22}. While defense encompasses a broader range of strategies and measures, repair serves as a specific action to fix or address vulnerabilities. The following section provides a detailed description of the defense methodologies involved.

\subsection{Contract Analyzer}
\label{sec:5_1}
Contract analyzers play a crucial role in reducing the risk of vulnerabilities in smart contracts before their deployment. Researchers employ various methodologies to analyze smart contracts, many of which are publicly available under open-source licenses. There are five common methodologies for smart contract analysis, such as formal verification, symbolic execution, fuzzing, intermediate representation, and machine learning.

\subsubsection{Formal Verification}
It is a mathematical-based technique to build and check formal proofs that satisfy a particular property. Formal verification is applied to ensure that software behaves and performs as expected in its specifications and requirements based on large reachable state spaces. Smart contracts are often written in programming languages that are amenable to formal verification \cite{harz2018towards, jiao2020semantic}. For example, the Solidity programming language used for Ethereum smart contracts has a well-defined syntax and semantics that can be precisely modeled and verified. For smart contracts, we distinguish mainly two families of formal verification methods, namely model checking and theorem proving.

\textbf{Model checking} It lists all possible states and checks them individually to confirm whether the contract has the corresponding characteristic. Compared to other programs, smart contracts have some unique characteristics that make model checking particularly well-suited for vulnerability detection \cite{garfatta2021survey, AbdellatifB18, AhrendtBEPPRS19, crincoli2022}. First, model checking is most effective on systems with a finite number of states, and smart contracts are often small enough to be modeled in their entirety. Second, smart contracts are often designed to be deterministic, which makes it easier to construct formal models to accurately capture their behavior. This technique is relevant to check partial specifications early in the design process \cite{NehaiPD18}. Model checkers, such as SIPN \cite{BaiCDH18}, FDR \cite{QuHCWML18}, and NuSMV \cite{NehaiPD18}, are utilized to successfully verify the correctness and necessary properties of smart contracts \cite{QuHCWML18}.  Model checking is typically used to detect specific types of vulnerabilities, such as buffer overflows or integer overflows.

\textbf{Theorem proving} It is a technique that describes the desired properties of a system using mathematical logic and uses a theorem prover to generate proofs that verify these properties based on evidence rules. In the context of smart contracts, theorem proving is used to ensure that a contract satisfies a specific set of properties, such as correctness, safety, or liveness. Since smart contracts are typically deterministic, theorem proving can construct mathematical proofs to demonstrate the satisfaction of these properties. 
Unlike model checking, which is limited to finite systems, theorem proving can handle the verification of infinite systems. This makes it well-suited for analyzing smart contracts that may involve complex interactions and potential infinite behaviors. 

Several theorem provers, such as Coq and Isabelle/HOL, have been developed to provide formal semantics and support the theorem-proving process for smart contracts. 
For instance, Amani et al. extended the existing definition of the Ethereum Virtual Machine (EVM) into Isabelle/HOL with the consideration of gas, allowing for formal verification of EVM-based smart contracts. However, it's worth noting that formal verification through theorem proving is a semi-automated process that often requires manual interaction. It is commonly used to detect broader classes of vulnerabilities, including logic errors and design flaws, rather than specific instances of vulnerabilities.

\subsubsection{Symbolic Execution}
It systematically explores more possible execution paths simultaneously to trigger deep program errors. This approach does not require a specified input value but abstracts the input values into symbols. 

From the vulnerability detection perspective, symbolic execution offers developers specific input to the triggered vulnerability, which can be used to confirm or debug. Symbolic execution has the advantage of achieving high test coverage with as few test cases as possible, thereby digging out deep program errors. 
Moreover, symbolic execution is often combined with constraint solving to reason whether an execution path is reachable. However, when the program is large and complex, symbolic execution will generate too many paths which may lead to path explosion. A smart contract can be a maximum of 24KB or it will run out of gas \cite{buterin2013ethereum}. Therefore, symbolic execution is perhaps the most popular approach for smart contracts. Moreover, Z3 SMT solvers are used to check which paths are flexible. 

Oyente \cite{LuuCOSH16} was the first attempt for smart contract verification, using a Control Flow Graph (CFG) representation of the bytecode to identify vulnerabilities like \textit{transaction-ordering dependence}, \textit{timestamp dependence}, \textit{mishandled exceptions} and \textit{reentrancy}. Other tools, like Manticore and Maian, have extended the capabilities of symbolic execution to detect additional vulnerabilities such as \textit{arithmetic overflow/underflow}, \textit{unsafe suicide},  and \textit{unchecked send}. 
However, scalability is a concern with symbolic execution, especially in complex application scenarios. The exploration of deep program paths can be slow and resource-intensive.

\subsubsection{Fuzzing}
It is a software testing technique that involves executing target programs with a large number of abnormal or random test cases to detect vulnerabilities. It has gained significant attention in both industry and academia \cite{LiJCLLCLWBCL021, HeBATV19} due to its simplicity and practical effectiveness in identifying software vulnerabilities. Major software vendors like Google  \cite{oss-fuzz} and Microsoft \cite{onefuzz} employ fuzzing techniques to uncover vulnerabilities in their products. 

In the context of smart contracts, fuzzing has been utilized as a means of vulnerability detection, although there are relatively fewer works specifically focused on smart contract fuzzing in recent years. ContractFuzzer \cite{0001LC18}, for example, uses the Application Binary Interface (ABI) specification of contracts as input for fuzzing to detect vulnerabilities. It relies on user-provided input seeds. Echidna \cite{echidna}, on the other hand, uses falsified user-defined predicates or Solidity assertions as input seeds, which are then subjected to grammar-based fuzzing to detect vulnerabilities. Harvey \cite{WustholzC20} incorporates a prediction component to generate new input seeds for gray box fuzzing. sFUZZ \cite{NguyenP0L020} adopts an adaptive strategy to select input seeds, which are then fed into the prominent fuzzer AFL (American Fuzzy Lop). However, these methods are more effective in finding shallow bugs and less effective in identifying bugs that lie deep in the execution flow. 

This limitation is due to the heavy reliance on input seeds in fuzzing. An alternative approach that has shown promising results in traditional programs is hybrid fuzzing, which combines fuzzing with symbolic execution. 
ILF (Imitation Learning Fuzzer)  \cite{HeBATV19} addresses this limitation by using a symbolic execution expert to generate a large number of training sequences, which are then fed into imitation learning prior to fuzzing. ILF achieves improved coverage and performs well on both large and small contracts. 
However, ILF is limited to the contracts used for imitation learning in its training phase. 
ConFuzzius \cite{TorresIGS21}, on the other hand, leverages lightweight symbolic execution to analyze execution traces and employs a constraint solver to obtain input seeds for the fuzzer. This approach enhances the effectiveness of fuzzing by incorporating symbolic execution-based analysis. 
By combining fuzzing with symbolic execution or other complementary techniques, researchers aim to enhance the effectiveness of vulnerability detection in smart contracts, addressing both shallow and deep vulnerabilities.

\subsubsection{Intermediate Representation}
To accurately analyze smart contracts, some researchers also explore converting contracts into an intermediate representation (IR) with highly semantic information, which is more suitable for the analysis and detection of common security issues. Different from formal verification, IR relies on semantic-based transformation. The analysis process can be divided into four stages: lexical analysis, syntax analysis, semantic analysis, and transformation.
The lexical analysis uses Scanner to check whether the input code is a combination of several legitimate words; the syntax analyzer checks whether the combination of these legitimate words meets grammatical rules; semantic analysis checks whether semantics are reasonable; the transformer converts source code or bytecode into machine code, such as XML. Then, the analyzer detects vulnerabilities through specific methods. 

Slither \cite{FeistGG19} transfer contracts to its internal representation language (SlithIR) which uses Static Single Assignment (SSA) form to facilitate the computation of code analyses. 
EthIR \cite{AlbertGLRS18} based on Oyente translates CFGs to a rule-based representation (RBR) of the bytecode.
Smartcheck \cite{TikhomirovVITMA18} directly translates source code into an XML-based IR and then checks it against XPath patterns.  
MadMax \cite{GrechKJBSS18} based on the Vandal \cite{abs-1809-03981} decompiler translates EVM bytecode to a structured IR to check gas-related Vulnerabilities.
NeuCheck \cite{LuWZSE21} employs the Solidity parser ANTLR to complete the transformation from source code to an IR (XML parse tree).

However, there are two challenges in IR analysis: (1) Because of semantic heterogeneity, it is unavoidable to produce semantic missing during the security analysis. (2) Compared with other analysis methodologies, IR takes more processing time.

\subsubsection{Machine Learning}
Machine learning demonstrated significant potential in program security, often outperforming traditional methods in various aspects \cite{ZhengGWLXLC20, ChakrabortyKDR22, hanif2021rise}. Unlike traditional methods, machine learning combines static analysis and dynamic detection. This combination addresses the high false negative rate of static analysis and the low code coverage of dynamic analysis. Furthermore, machine learning exhibits excellent scalability and adaptability to novel vulnerabilities.

Several prior works have employed machine learning techniques to analyze smart contracts. Tann et al. \cite{abs-1811-06632} introduced a long short-term memory (LSTM) \cite{ZhengGWLXLC20} model to handle the semantic representations of smart contract opcode to detect contract security threats. Their model can achieve higher detection accuracy than symbolic execution analyzer Maian \cite{NikolicKSSH18}, where both are based on the same vulnerabilities taxonomy. 
Qian et al. \cite{QianLHZW20} applied a bi-directional long short-term memory with attention mechanism (BLSTM-ATT) in their sequential model to \textit{reentrancy} detection. This framework converts source code into contract snippets and feeds the sequential model with feature vector representations parsed from these snippets.  

Zhuang et al. \cite{ZhuangLQLWH20} proposed a degree-free graph convolutional neural network (DR-GCN) and a novel temporal message propagation network (TMP) for vulnerability detection. In their approach, the source code of a contract is converted to a contracted graph, which is then normalized through a node elimination process. The normalized graphs are fed to DR-GCN and TMP for vulnerability modeling and detection.
ContractWard \cite{WangSXLWS21} trains its machine learning model using bigram features extracted from the opcodes of the compiled smart contract.

SmartMixModel \cite{ShakyaMHMC22} extracts high-level syntactic features from source code as well as low-level bytecode features from the smart contract. And then these features are trained on a machine learning model and deep learning model to detect vulnerabilities. 
ESCORT \cite{sendner2023smarter} introduces a deep learning-based method for vulnerability detection in smart contracts, combined with distinct branches for learning specific features of various vulnerability types. When a new vulnerability type is identified, ESCORT seamlessly integrates a new branch into the existing feature extractor and trains it with minimal data.

In addition to these methods, some researchers suggest that combining machine learning with fuzzing could enhance detection efficiency. 
SoliAudit \cite{LiaoTHT19} employs machine learning to detect known vulnerabilities without the need for expert knowledge or predefined patterns, while fuzzing is used to identify potential weaknesses. Although there is no direct correlation between the two methods, fuzzing complements machine learning in vulnerability detection. ILF \cite{HeBATV19} utilizes an imitation learning model to develop a fuzzer from training sequences, demonstrating another innovative approach to enhancing vulnerability detection through machine learning and fuzzing.

Recent studies have explored the potential of large language models (LLMs) in enhancing software security, leveraging their capabilities in code comprehension, generation, and analysis. Researches have demonstrated the effectiveness of LLMs in identifying vulnerabilities, verifying compliance, and assessing logical correctness. Chen et al. \cite{chen2023diversevul} demonstrated that LLMs (GPT-2, T5) trained on a high-quality dataset of 18,945 vulnerable C/C++ functions outperformed other machine learning methods, including Graph Neural Networks, in vulnerability prediction.

The application of LLMs to smart contract security has gained significant attention. David et al. \cite{david2023you} investigated the utility of LLMs, like GPT-4 and Claude, in conducting security audits of DeFi smart contracts. Chen et al. \cite{chen2023chatgpt} conducted a comparative analysis of GPT’s performance in identifying smart contract vulnerabilities against other established tools, utilizing a publicly accessible dataset. Sun et al. \cite{sun2023gpt} tested GPT’s ability to match candidate vulnerabilities using a binary response format, where GPT responds with ‘Yes’ or ‘No’ to potential matches with predefined scenarios. They also highlighted potential false positives due to GPT’s inherent limitations. Shou et al. \cite{shou2024llm4fuzz} integrated the Llama-2 model into the fuzzing process to detect vulnerabilities in smart contracts, aiming to address inefficiencies in traditional fuzzing methods. Sun et al. \cite{sun2024llm4vuln} compared open-source language models like Mixtral and CodeLlama against GPT-4 for smart contract vulnerability detection. They found that GPT-4, leveraging its advanced Assistants' functionalities, significantly outperformed open-source alternatives.

\subsection{Security Enhancement}
Before deployment, some measures can be done to enforce the security of smart contracts. Gas and data are the most important factors for smart contracts. Thus, if gas cost and data privacy have been carefully checked before being deployed, the security of smart contracts will be enhanced. There are some methods and tools for gas cost and data privacy.

\subsubsection{Gas Estimation and Optimization}
Gas is one of the most important mechanisms in EVM for assigning a cost to the execution of an instruction. This mechanism can effectively prevent resource abuse (especially DoS attack) and avoid ``infinite" loops \cite{ChenLWCLLAZ17}. When issuing a transaction, the sender needs to specify a gas limit and a gas price before submitting it to the network. 

Gas represents much more than just the cost of processing transactions on the Ethereum network. It enables smart contracts to run various applications, forming a decentralized web. Thus, while gas could technically be described as "transaction fees," this term should be used with caution. Several gas-related vulnerabilities exist, as discussed in Section \ref{vulnerability}. If a transaction runs out of gas during execution, the EVM throws an exception, immediately reverting to the state before execution and consuming all the gas provided by the sender.

To prevent running out of gas, many Ethereum wallets, such as Metamask \cite{METAMASK}, can statically estimate the cost of a transaction before it is executed.
However, there are many operations that are difficult to estimate during executions. In order to address this problem, some researchers proposed some methods that estimate gas and optimize smart contracts \cite{AlbertGRS19,ashraf2020gasfuzzer, danielius2020}. 

Albert et al. \cite{AlbertGRS19} proposed an automatic gas analyzer Gastap that infers gas upper bounds for all its public functions. Gastap requires complex transformation and analysis of the source code that includes several key techniques, including Oyente \cite{LuuCOSH16}, EthIR \cite{AlbertGLRS18}, Saco \cite{AlbertAFGGMPR14} and Pubs \cite{AlbertAGP08}. Gasol \cite{AlbertCGRR20}, an extension of Gastap, introduces an automatic optimization of the selected function that reduces inefficient gas consumption. 

Li et al. \cite{ChenFLZLLXCZ21} developed a GasChecker tool that identifies the gas-inefficient code of smart contracts. They summarized ten gas-inefficient programming patterns that assisted users in better tailoring contracts to avoid gas waste. Gas Gauge \cite{abs-2112-14771} can automatically estimate the gas cost for the target function and the loop-bound threshold. Besides, Gas Gauge can find all the loops and furnish the gas-related instances to help developers with suggestions. Li et al. \cite{LiNCYH20} estimated the gas for new transactions by learning the relationship between historical transaction traces and their gas costs.

\subsubsection{Data Privacy-Preserving}
In addition to gas concerns, another major concern for the smart contract is private data \cite{KalodnerGCWF18}. Since the openness and transparency of public blockchains, the privacy overlay feature on the chain is absent. This not only leads to some security issues but also prevents their wider adoption. 
It is problematic for applications that handle sensitive data such as voting schemes \cite{McCorrySH17}, electronic medical records\cite{LiuLYZDG18}, or crowdsensing \cite{PerezZ22}. 

A promising approach to handling private data involves designing new blockchain infrastructures with built-in privacy support \cite{KosbaMSWP16}. Several proposed blockchain infrastructures, including Hawk \cite{KosbaMSWP16}, Arbitrum \cite{KalodnerGCWF18}, Ekiden \cite{ChengZKHHJJ0S19}, and Town Crier \cite{zhang2016town}, support private data through trusted third-party mechanisms.
Hawk \cite{KosbaMSWP16} enables programmers to write private smart contracts without specialized knowledge and generates efficient cryptographic protocols for inter-party interactions. Arbitrum \cite{KalodnerGCWF18} facilitates private smart contract creation by emulating a virtual machine (VM) and enables honest parties to update the VM state on-chain. Both Hawk and Arbitrum utilize trusted managers, implementable through trusted computing hardware or multi-party computation among users.
Ekiden \cite{ChengZKHHJJ0S19} processes smart contracts with private data off-chain in trusted execution environments (TEEs), while ensuring secure on-chain interactions between contracts.

An alternative approach involves using cryptographic primitives, particularly zero-knowledge proofs. Hawk \cite{KosbaMSWP16} employs zero-knowledge proofs to ensure correct contract execution and maintain money conservation on-chain. Steffen et al. introduced zkay contracts, which incorporate private data protected by non-interactive zero-knowledge (NIZK) proofs \cite{SteffenBGMTV19}. For enhanced blockchain executability, zkay contracts are transformed into equivalent contracts that maintain privacy and functionality. Zapper \cite{SteffenBV22} utilizes NIZK proofs for private state updates and employs an oblivious Merkle tree to conceal sender and receiver identities.

\subsection{Runtime Monitoring}
The capability to deploy smart contracts is one of the most important features of blockchains. Once deployed on a blockchain, smart contracts become immutable, including any vulnerabilities they may have. While tools are available for scanning smart contracts prior to deployment, they often have limited scopes and may not detect all vulnerabilities, leaving room for unknown runtime attacks. As a solution to enhance the security of post-deployment smart contracts, runtime monitoring techniques have emerged. These techniques analyze and monitor the runtime behavior of smart contracts, providing higher coverage and precision in detecting attacks compared to pre-deployment analysis alone \cite{ellul2018runtime}. Runtime monitoring leverages real-time information during the execution of smart contracts, enabling the detection of vulnerabilities and malicious activities that may occur at runtime. It can be categorized into two main types based on the detection target: transaction monitoring and state monitoring. 

\subsubsection{Transactions Monitoring}
As Ethereum can be seen as a transaction-based state machine, a transaction contains much information aiming to change the blockchain state.  This transaction information can also be used to detect and prevent attacks.

ECFChecker \cite{GrossmanAGMRSZ18} is the first dynamic detection tool designed to identify transactions that include reentrancy vulnerabilities. It examines transactions to determine if they exhibit the characteristics of a reentrancy attack, where a contract can be called recursively before previous invocations have been completed.
Sereum \cite{RodlerLKD19}, on the other hand, aims to prevent reentrancy attacks by employing taint tracking techniques. It tracks the flow of data from storage variables to control-flow decisions, helping identify potential vulnerabilities \cite{hu2021transaction, 0002CLLGZLZCHTL20}. Both ECFChecker and Sereum rely on modified versions of the Ethereum Virtual Machine and primarily focus on detecting reentrancy attacks.

In contrast, ÆGIS \cite{TorresBNJ19} takes a broader approach by providing an extensible framework for detecting new vulnerabilities in smart contracts. It maintains attack patterns and reverts transactions that match these patterns, thereby enhancing security. The framework allows for the storage and voting of new attack patterns through a smart contract, enabling the community to actively contribute to the detection and prevention of novel attacks. 

SODA \cite{0002CLLGZLZCHTL20}, is based on a modified EVM-based client and provides a platform for developing various applications to detect malicious transactions in real time. It offers flexibility and extensibility in creating online apps for monitoring and identifying potentially harmful transactions. SODA has been integrated into popular blockchains that support the EVM, increasing its accessibility and usability.
Horus \cite{ZhouQTLG21} leverages transaction graph-based analysis to identify the flow of stolen assets. By examining the transaction graph, Horus can trace the movement of assets and detect potential theft or unauthorized transfers.

\subsubsection{State Monitoring}
In addition to transactions, blockchain state variables can provide valuable information about the real-time status of a smart contract. Monitoring and analyzing the state variables can be an effective approach to detect and prevent attacks.

Solythesis \cite{LiCL20} allows users to instrument user-specified invariants into smart contract code. These invariants represent specific conditions that should always hold true during the execution of the contract. By tracking the transactions that violate these invariants, Solythesis can efficiently enforce powerful monitoring of the contract's state. This approach helps identify abnormal or unexpected changes in the state variables and allows for proactive detection of potential vulnerabilities.

ContractGuard \cite{WangHXZC20} takes a similar approach but introduces the concept of an intrusion detection system (IDS) to monitor the behavior of a deployed smart contract. The IDS continuously monitors the state variables and looks for abnormal or suspicious behaviors that deviate from expected patterns. If an abnormal state is detected, ContractGuard can roll back all the changes to the contract state and notify the relevant users about the potential intrusion.

\subsection{Post-deployment Repair}
Ideally, smart contracts should be deployed with the highest possible level of security. However, this presents a challenge not only for smart contracts but also for all software programs. Furthermore, the underlying blockchain technology ensures immutability, meaning that the past cannot be altered. Consequently, updating the code of a deployed contract or addressing vulnerabilities becomes unfeasible. Apart from fixing vulnerabilities, there are numerous other reasons to modify contract code, including adapting business logic or enhancing functionality.

\subsubsection{Contract Migration}
In many cases, eploying a new instance of the contract and migrating the old contract's data to it is a viable approach. This process, known as contract migration, has been successfully implemented in various token migration events, such as Augur, VeChainThor, and TRON.

A typical contract migration process involves two main steps: data recovery and data writing \cite{josselinfeist}. During data recovery, data from the old contract is extracted, including both public and private variables. While retrieving public variables is relatively straightforward, handling private variables and mappings can be more complex, requiring special measures to ensure data integrity and consistency.

To address the challenges of data migration, some developers use separate contracts to store and manage data \cite{ProxyPatterns}. This approach simplifies the migration process by separating the logic from data storage. By migrating the logic to a new contract while keeping the data in a separate contract, the migration becomes more manageable.
It's important to note that contract migration procedures can be costly, especially for token-based contracts with a large number of accounts \cite{josselinfeist}. Migrating data for many accounts requires careful planning and execution to avoid potential issues and ensure the accuracy and security of the migrated data.

\begin{figure}[t]
	\centering
	\includegraphics[width=4.8in]{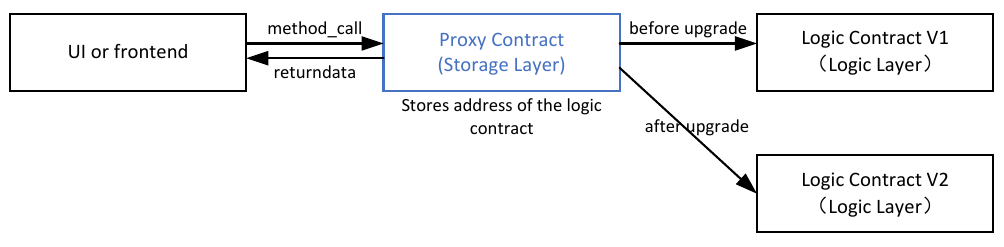}
	\caption{Contract upgrade with proxy}
	\label{fig_proxypatterns}
\end{figure}

\subsubsection{Upgradable Smart Contracts}
Upgradable contracts offer another promising approach to modifying smart contract code \cite{josselinfeist2}. This method incorporates an upgradability mechanism into smart contracts, enabling seamless upgrades without the need for data migration or updating external references.

An upgradable smart contract can be implemented using a proxy pattern architecture. In this architecture, a single contract is divided into two separate contracts: one for logic execution and another for data storage, as illustrated in Figure \ref{fig_proxypatterns}. The logic contract does not store any state but implements all business logic, while the proxy contract holds all funds and internal states without implementing any business logic. This separation allows for quick and cost-effective upgrades by modifying or replacing only the logic contract, while the proxy contract remains intact, maintaining all existing data and external references. This ensures seamless upgrades without disrupting the contract's functionality.

Several approaches have been proposed for implementing upgradable contracts. Zeppelin \cite{ProxyPatterns} introduced three proxy pattern architectures: inherited storage, eternal storage, and unstructured storage. These patterns are collectively known as delegate-proxy patterns because they rely on the DELEGATECALL instruction. Based on this concept, EVMPatch \cite{RodlerLKD21} uses a proxy pattern to facilitate quick and cost-effective smart contract upgrades. It provides a framework for patching vulnerabilities or adding new features to deployed contracts without requiring data migration or disrupting the contract's functionality. 

Despite their benefits, upgradable smart contracts pose notable security challenges. They necessitate extensive security knowledge for secure implementation and can be exploited for malicious purposes. To solve this issue, Bodell et al. \cite{bodell2023proxy} develop a complete taxonomy to comprehensively characterize the unique behaviors of upgradable smart contracts. 

\begin{figure}[t]
    \centering
    \includegraphics[width=3in]{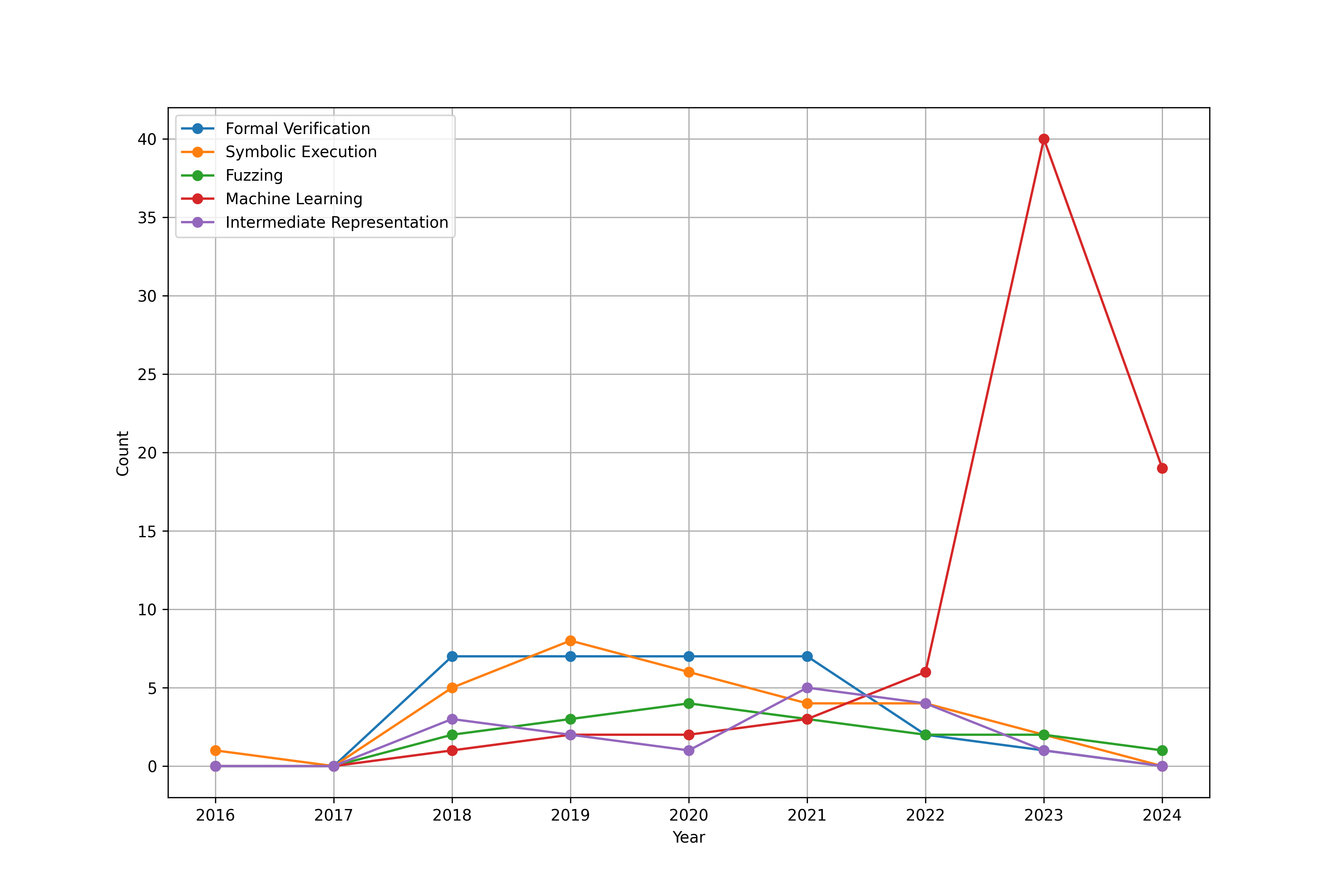}
    \caption{Trends in Method Usage Over the Years}
    \label{fig: method_trend}
\end{figure}

\subsection{Evolving Trends in Contract Analyzer Usage}
Smart contract analyzers, as the primary strategy for identifying vulnerabilities in smart contracts, have been extensively studied by researchers. As discussed in Section \ref{sec:5_1}, there are five common methodologies for smart contract analysis. The utilization of these techniques has varied significantly over time.

Figure \ref{fig: method_trend} illustrates a significant shift in method usage trends after 2021. Before 2021, methods like Formal Verification, Symbolic Execution, and Fuzzing showed steady or increasing usage, peaking between 2018 and 2020. Post-2021, however, these methods experienced a notable decline, indicating a decrease in their adoption or relevance. Formal Verification and Symbolic Execution, once widely adopted, showed a marked decrease, suggesting a shift in research focus or industry practices. Fuzzing exhibited a similar trend, peaking around 2020 before declining.
Intermediate Representation (IR), despite its relatively stable usage, also declined post-2021, reinforcing the general downward trend across most methods. 

On the other hand, Machine Learning stands out as an exception to this trend. While Machine Learning had fluctuating usage in earlier years, its relevance surged dramatically in 2023, reflecting its growing importance and widespread application in contemporary fields. The boost of Machine Learning also reveals that more and more researchers are paying attention to smart contract security, leveraging advanced techniques to address emerging challenges.

This section has provided a comprehensive examination and categorization of defense methodologies designed to prevent or mitigate smart contract attacks. These findings offer a comprehensive response to {RQ3}, as defined in Section \ref{survey_methodology}. However, there is a notable lack of empirical evidence to evaluate the effectiveness of these methodologies and their associated tools. Despite claims of superiority, empirical studies are necessary to determine which methods are practical and beneficial in real-world applications.

\section{Evaluation}
\label{evaluation}
This study aims to provide a comprehensive overview of SOTA automated analysis tools available for smart contracts. Given that the field of smart contract analysis is relatively new and rapidly evolving, it can be challenging to stay updated with the latest developments and understand the strengths and limitations of existing tools. Therefore, we have conducted an extensive review of the literature and websites to compile a list of the most promising analysis tools for smart contracts. To address RQ4 and RQ5, which focus on evaluating the performance of existing tools, we propose a systematic evaluation approach that involves selecting appropriate tools, utilizing datasets of smart contracts, defining criteria for the assessment, and conducting experiments.

\subsection{Experimental Settings}
\subsubsection {Tool Selection} 
Vulnerability detection tools are widely employed to assist developers in identifying vulnerabilities within smart contracts. Numerous analysis tools has been developed for this purpose. We have compiled a list of 169 such tools based on academic literature and online sources. The compiled list, including key properties such as publication venue, methodology, input type, open-source repository, and vulnerability identifier, is available in a comprehensive table\footnote{\url{https://github.com/WeiZ-boot/survey-on-smart-contract-vulnerability}}. 

For tool selection in our comparative analysis, we applied four criteria:

\begin{itemize}[leftmargin=*]
  \item {Available}: The tool must be publicly available and accessible for download or running with a command line interface (CLI).
  \item {Functionality}: The tool must be designed for smart contracts and capable of detecting vulnerabilities.  This excludes tools that only construct artifacts like control flow graphs.
  \item {Compatibility}: The tool must operate on the source code of the smart contract, excluding those that only consider EVM bytecode.
  \item {Documentation}: The tool must provide comprehensive documentation and user guides.
\end{itemize}

Application of these criteria to our comprehensive tool list resulted in the identification of 14 tools meeting all requirements: ConFuzzius \cite{TorresIGS21}, Conkas \cite{conkas}, Maian \cite{NikolicKSSH18}, Manticore \cite{manticore}, Mythril \cite{DurieuxFAC20}, Osiris \cite{TorresSS18}, Oyente \cite{LuuCOSH16}, Securify \cite{BrentGLSS20}, sfuzz \cite{NguyenP0L020}, Slither \cite{slither}, Smartcheck \cite{TikhomirovVITMA18}, solhint \cite{solhint}, GPT-4o \cite{achiam2023gpt}, and Meta-Llama-3.1-8b \cite{touvron2023llama}.

\subsubsection{Benchmarking Dataset}
A key issue when evaluating analysis tools is obtaining a sufficient number of vulnerable smart contracts. Despite the availability of numerous open-source analysis tools, comparing and reproducing results can be challenging due to the lack of publicly available datasets. Most analysis tools primarily check for only some of the well-known Ethereum smart contract vulnerabilities. To evaluate the effectiveness of any analysis tool, establishing a standard benchmark is crucial. While several researchers have published their datasets \cite{DurieuxFAC20, GhalebP20, TorresIGS21}, these often have limitations such as small sample sizes or an uneven distribution of vulnerable contracts.

To address these limitations, we have compiled an extensive, annotated dataset comprising 110 distinct smart contract test cases. These cases are categorized into 11 sub-datasets: 10 sub-datasets contain known vulnerabilities corresponding to the top 10 vulnerability categories mentioned in Section \ref{vulnerability}, and one sub-dataset represents safe contracts.

The selection of these contracts was carefully curated to cover a broad spectrum of code complexities and application scenarios, ensuring a comprehensive benchmark for tool evaluation. Our benchmark dataset is publicly available on GitHub for community use and further research development \footnote{\url{https://github.com/WeiZ-boot/Smartcontract-benchmark}}.

\subsubsection{Evaluation Criteria}
To systematically and objectively assess the quality of selected smart contract analysis tools, we developed a set of evaluation criteria based on the internationally recognized ISO/IEC 25010 standard. This standard offers a comprehensive framework for software product quality assessment, focusing on eight key characteristics. For our study, we tailored these characteristics to the specific requirements of smart contract analysis, concentrating on four of the most relevant aspects: detection suitability (functional suitability), resource Efficiency (performance efficiency), version compatibility (compatibility), and category coverage (usability).

\begin{itemize}[leftmargin=*]
  \item \textbf{Detection Suitability}: This criterion evaluates whether the analysis tools meet the functional requirements specified by users, primarily focusing on their ability to detect vulnerabilities in smart contracts. It assesses the tools' capability to identify potential security issues within contract code. The suitability of detection encompasses various aspects: accuracy, precision, recall, and adaptability.
  \item \textbf{Resource Efficiency}: This aspect assesses whether a software product can effectively make use of given resources. In the context of our study, this is quantified by examining the execution time required by various smart contract analysis tools.  Efficient performance is crucial for practical applicability, especially when dealing with large volumes of contracts.
  \item \textbf{Version Compatibility}: This criterion concerns whether a software product can consistently perform its functions while exchanging. In our study, we examines the tools' ability to function consistently across different versions of the Solidity programming language. Given the frequent updates and changes in Solidity, high compatibility ensures broader applicability and longevity of the tools.
  \item \textbf{Category Coverage}: This aspect evaluates whether users can effectively and efficiently use a software product to complete tasks. In our study, we measure the tools' ability to cover a wide range of vulnerability types. A tool with extensive category coverage is more useful to users, as it enables the detection of various vulnerabilities within a contract, thereby enhancing overall usability.
\end{itemize}

To further elaborate, detection suitability can be seen as a binary classification problem. The primary objective of the assessment tool is to accurately determine the presence or absence of specific vulnerabilities in a smart contract. 
This binary classification method simplifies the evaluation methodology and provides an effective measure of the tool's precision in vulnerability identification. 
The classification outcomes are categorized into four distinct groups: 

\begin{itemize}[leftmargin=*]
  \item \textbf{True Positive (TP)}: Tool correctly identifies a vulnerability in a contract when it actually exists.
  \item \textbf{False Positive (FP)}: Tool incorrectly identifies a vulnerability in a contract when none exist.
  \item \textbf{False Negative (FN)}: Tool fails to identify a vulnerability when one actually exists.
  \item \textbf{True Negative (TN)}: Tool correctly identifies that a contract does not have a vulnerability when it does not.
\end{itemize}

To evaluate tool's detection effectiveness, we employ three key metrics: Precision, which is the ratio of true positive results to all positive results predicted by the tool (Precision = TP / (TP + FP)); Recall rates, which is the ratio of true positive results to all actual positive cases (Recall = TP / (TP + FN)); and F1-score, which is the harmonic mean of Precision and Recall (F1 = (2 * Precision * Recall) / (Precision + Recall)).  

\subsubsection{Hardware Configuration}
We obtained the most recent versions of the selected analysis tools from their respective public GitHub repositories, except for version 0.3.4 of the Manticore tool. The tools were executed on an Aliyun-hosted Ubuntu 22.04 LTS machine, configured with an Intel(R) Core(TM) i5-13400 CPU and 32GB of RAM. Additionally, for language model validation, we maintained consistency by using the same NVIDIA A10 GPUs with 24GB of VRAM.

\subsection{Experimental Results}

\begin{table}[t]
\centering
\renewcommand{\arraystretch}{1.1}
\footnotesize
\setlength{\tabcolsep}{3pt}
\caption{Comparative Analysis of Vulnerability Detection Accuracy Across Smart Contract Detection Tools}
\label{tab:test-TP}
\begin{tabular}{@{}l*{14}{c}@{}}
\toprule
\textbf{Category} & 
\rotatebox{90}{\textbf{ConFuzzius}} & 
\rotatebox{90}{\textbf{Conkas}} & 
\rotatebox{90}{\textbf{Maian}} & 
\rotatebox{90}{\textbf{Manticore}} & 
\rotatebox{90}{\textbf{Mythril}} & 
\rotatebox{90}{\textbf{Osiris}} & 
\rotatebox{90}{\textbf{Oyente}} & 
\rotatebox{90}{\textbf{Securify}} & 
\rotatebox{90}{\textbf{sFuzz}} & 
\rotatebox{90}{\textbf{Slither}} & 
\rotatebox{90}{\textbf{Smartcheck}} & 
\rotatebox{90}{\textbf{solhint}} &
\rotatebox{90}{\textbf{GPT-4o}} & 
\rotatebox{90}{\textbf{Llama-3.1-8b}}\\
\midrule
Reentrancy & 9 & 10 & 0 & 8 & 9 & 7 & 7 & 8 & 6 & 9 & 8 & 0 & 10 & 9 \\
Arithmetic & 7 & 9 & 0 & 7 & 7 & 9 & 9 & 0 & 6 & 0 & 1 & 0 & 10 & 10 \\
Gasless Send & 0 & 0 & 0 & 0 & 0 & 0 & 0 & 0 & 3 & 0 & 7 & 0 & 5 & 2 \\
Unsafe Suicide & 4 & 0 & 4 & 0 & 3 & 0 & 0 & 0 & 0 & 6 & 0 & 4 & 9 & 9 \\
Unsafe Delegatecall & 1 & 0 & 0 & 4 & 6 & 0 & 0 & 0 & 5 & 7 & 0 & 0 & 9 & 10 \\
Unchecked Send & 9 & 10 & 0 & 2 & 9 & 0 & 0 & 0 & 0 & 0 & 8 & 6 & 8 & 6 \\
TOD & 2 & 7 & 0 & 0 & 0 & 0 & 2 & 2 & 0 & 0 & 9 & 0 & 3 & 3 \\
Timestamp Manipulation & 8 & 8 & 0 & 7 & 6 & 2 & 0 & 0 & 1 & 8 & 0 & 10 & 10 & 7 \\
Authorization through tx.origin & 0 & 0 & 0 & 3 & 6 & 0 & 0 & 0 & 0 & 0 & 0 & 9 & 10 & 6 \\
Bad Randomness & 2 & 0 & 0 & 0 & 8 & 0 & 0 & 0 & 0 & 0 & 0 & 5 & 10 & 2 \\
\midrule
\textbf{Total} & 42 & 44 & 40 & 31 & 54 & 18 & 18 & 10 & 21 & 30 & 33 & 34 & \textbf{84} & 64 \\
\bottomrule
\end{tabular}
\end{table}

\subsubsection{Detection Suitability}
We first measure the detection effectiveness of the selected tools in identifying vulnerabilities. We test the selected tools on our benchmark, and the results are summarized in Table \ref{tab:test-TP}. This table presents an overview of the strengths and weaknesses of the selected tools across the common 10 categories discussed in Section \ref{vulnerability}. The numbers in each cell show the number of TPs identified by each tool for each vulnerability category.

GPT-4o shows the highest overall performance, detecting 84 out of 100 vulnerabilities across all categories. Llama-3.1-8b is the second-best performer, detecting 64 vulnerabilities. Among traditional tools, Mythril performs best with 54 TPs. The results also reveals that different tools perform better at identifying certain vulnerability categories. For example, GPT-4o, Conkas, and Llama-3.1-8b perform best at reentrancy vulnerabilities, while Smartcheck performs best at Gasless Send vulnerabilities. LLMs (GPT-4o and Llama-3.1-8b) generally outperform traditional tools across most categories.

To further study the tools' effectiveness, we add safe contracts to our experiments. The results are presented in Table \ref{tab:test-accuracy}. GPT-4o shows the best overall performance with the highest Accuracy (79\%) and F1-score (0.88). Llama-3.1-8b and Mythril tie for second place in Accuracy (58\%), but Llama-3.1-8b has a higher F1-score (0.74 vs 0.70). Interestingly, Llama-3.1-8b has the highest false positive (FP) count (10), followed by GPT-4o (7). While LLMs (especially GPT-4o) show promising results in overall vulnerability detection, they may exhibit high FP rates. Traditional tools, despite lower overall performance, might be valuable for their high precision in specific scenarios.

These findings suggest that while LLM-based tools show great promise in enhancing smart contract security, they should be used in conjunction with traditional tools and expert review for optimal results. 
\begin{table}[t]
\centering
\renewcommand{\arraystretch}{1.1}
\footnotesize
\setlength{\tabcolsep}{3pt}
\caption{Accuracy and F1-score Across Smart Contract Detection Tools}
\label{tab:test-accuracy}
\begin{tabular}{@{}l*{14}{c}@{}}
\toprule
\textbf{Metric} & 
\rotatebox{90}{\textbf{ConFuzzius}} & 
\rotatebox{90}{\textbf{Conkas}} & 
\rotatebox{90}{\textbf{Maian}} & 
\rotatebox{90}{\textbf{Manticore}} & 
\rotatebox{90}{\textbf{Mythril}} & 
\rotatebox{90}{\textbf{Osiris}} & 
\rotatebox{90}{\textbf{Oyente}} & 
\rotatebox{90}{\textbf{Securify}} & 
\rotatebox{90}{\textbf{sFuzz}} & 
\rotatebox{90}{\textbf{Slither}} & 
\rotatebox{90}{\textbf{Smartcheck}} & 
\rotatebox{90}{\textbf{solhint}} &
\rotatebox{90}{\textbf{GPT-4o}} & 
\rotatebox{90}{\textbf{Llama-3.1-8b}}\\
\midrule
TP & 42 & 44 & 4 & 31 & {54} & 18 & 18 & 10 & 21 & 30 & 33 & 34 & 84 & 64 \\
FN & 58 & 56 & 96 & 69 & 46 & 82 & 82 & 90 & 79 & 70 & 67 & 66 & {16} & 36 \\
FP & 3 & 4 & 0 & 0 & 0 & 2 & 2 & 0 & 1 & 0 & 0 & 0 & 7 & 10 \\
TN & 7 & 6 & 10 & 10 & 10 & 8 & 8 & 10 & 9 & 10 & 10 & 10 & 3 & 0 \\
\midrule
Accuracy & 45\% & 46\% & 13\% & 37\% & 58\% & 24\% & 24\% & 18\% & 27\% & 36\% & 39\% & 40\% & \textbf{79\%} & 58\% \\
Recall & 0.42 & 0.44 & 0.04 & 0.31 & 0.54 & 0.18 & 0.18 & 0.1 & 0.21 & 0.3 & 0.33 & 0.34 & 0.84 & 0.64 \\
Precision & 0.93 & 0.92 & 1 & 1 & 1 & 0.9 & 0.9 & 1 & 0.95 & 1 & 1 & 1 & 0.92 & 0.86 \\
F1-score & 0.58 & 0.59 & 0.08 & 0.47 & 0.70 & 0.3& 0.3 & 0.18 & 0.34 & 0.46 & 0.49 & 0.51 & \textbf{0.88} & 0.74 \\
\bottomrule
\end{tabular}
\end{table}
\subsubsection{Resource Efficiency}
Execution time is a crucial factor in evaluating a tool's effectiveness, as it directly impacts efficiency. We assessed the resource efficiency of each tool by measuring their execution time, as indicated in Table \ref{tab:test-performance}.

Our analysis revealed significant variations in execution time across different tools. Slither and solhint emerged as the fastest tools, with an average execution time of 1 second. GPT-4o also demonstrated high efficiency, with an average execution time of 2.2 seconds. On the other end of the spectrum, Conkas was the slowest, taking 48 minutes on average to complete its analysis. ConFuzzius and sFuzz also exhibited long execution time, which could be attributed to their utilization of fuzzing methods. 

In general, we observed that static analysis tools (such as Slither and solhint) and LLMs (like GPT-4o) are significantly faster than tools employing more complex analysis techniques.

\begin{table}[ht]
\centering
\renewcommand{\arraystretch}{1.1}
\footnotesize
\setlength{\tabcolsep}{3pt}
\caption{Execution Time, Compatibility Version ($S_v$) and Category Coverage ($S_c$) Scores for Each Tool}
\label{tab:test-performance}
\begin{tabular}{@{}lc*{13}{c}@{}}
\toprule
\textbf{Metric} & 
\rotatebox{90}{\textbf{ConFuzzius}} & 
\rotatebox{90}{\textbf{Conkas}} & 
\rotatebox{90}{\textbf{Maian}} & 
\rotatebox{90}{\textbf{Manticore}} & 
\rotatebox{90}{\textbf{Mythril}} & 
\rotatebox{90}{\textbf{Osiris}} & 
\rotatebox{90}{\textbf{Oyente}} & 
\rotatebox{90}{\textbf{Securify}} & 
\rotatebox{90}{\textbf{sFuzz}} & 
\rotatebox{90}{\textbf{Slither}} & 
\rotatebox{90}{\textbf{Smartcheck}} & 
\rotatebox{90}{\textbf{solhint}}  &
\rotatebox{90}{\textbf{GPT-4o}} & 
\rotatebox{90}{\textbf{Llama-3.1-8b}} \\
\midrule
Comp. ver. & 0.8.x & 0.5.x & 0.8.x & 0.8.x & 0.8.x & 0.4.21 & 0.4.19 & 0.5.11 & 0.4.24 & 0.6.x & 0.8.x & 0.8.x  & 0.8.x & 0.8.x \\
$S_v$ & {5} & 2 & {5} & {5} & {5} & 1 & 1 & 2 & 1 & 3 & {5} & {5} & 5 & 5 \\
$S_c$ & {8} & 5 & 1 & 6 & 7 & 3 & 3 & 2 & 5 & 4 & 5 & 5 & 10 & 10\\
Avg. Time & 18m9s & 48m & 57s & 11m2s & 4m17s & 1m32s & 4s & 27s & 18m & \textbf{1s} & 3s & \textbf{1s} & 2.2s & 47s \\
\bottomrule
\end{tabular}
\end{table}

\subsubsection{Version Compatibility and Category Coverage}
Given Solidity's continuous updates, it's essential for analysis tools to keep pace with the latest versions to ensure accurate vulnerability detection. We assessed each tool's compatibility with Solidity versions up to 0.8.19 (the latest version as of January 2023). We assigned a rating value ($S_v$) to different Solidity versions based on their compatibility with each tool. The rating value starts from 1 for version 0.4.x and increases to 5 for version 0.8.x.

The range of vulnerability categories a tool can detect is crucial in evaluating its effectiveness. Tools with wider category coverage may be more versatile and effective in comprehensive smart contract audits. We assigned a score ($S_c$) to each tool based on the number of vulnerability categories it can detect. For instance, a tool capable of detecting 5 categories of vulnerabilities receives an $S_c$ score of 5.

Table \ref{tab:test-performance} summarizes the $S_v$ and $S_c$ scores for each evaluated tool. Most tools, including ConFuzzius, Maian, Manticore, Mythril, Smartcheck, solhint, GPT-4o, and Llama-3.1-8b, support Solidity version 0.8.x, achieving the highest $S_v$ score of 5. GPT-4o and Llama-3.1-8b demonstrated the highest $S_c$ score of 10, indicating they cover all identified vulnerability categories.

\subsubsection{Overall Effectiveness}
To provide a comprehensive assessment of the smart contract analysis tools, we developed a weighted sum method that balances the key factors evaluated in the previous sections: detection effectiveness, resource efficiency, version compatibility, category coverage. We assign weights to each factor based on their relative importance. $\alpha$ to accuracy, $\beta$ to average execution time, $\gamma$ to compatibility version, and $ \delta$ to category coverage, where $\alpha + \beta + \gamma + \delta = 1$. 

The overall score for each tool is calculated using the following formula:

\begin{align*}
  Score = \alpha \times A *10  + \beta \times (1/AEX)*10 + \gamma \times S_v + \delta \times S_c
\end{align*}
where  $A$ is the accuracy value and $AEX$ is the average execution time.

\begin{table}[t]
\centering
\renewcommand{\arraystretch}{1.2}
\footnotesize
\setlength{\tabcolsep}{3pt}
\caption{Overall Scores for Each Tool with Different Weight Configurations}
\label{tab:test-score}
\begin{tabular}{@{}c|cccccccccccccc@{}}
\toprule
Weights: $\alpha$, $\beta$, $\gamma$, $\delta$&
\rotatebox{90}{\textbf{ConFuzz}} & 
\rotatebox{90}{\textbf{Conkas}} & 
\rotatebox{90}{\textbf{Maian}} & 
\rotatebox{90}{\textbf{Mantic}} & 
\rotatebox{90}{\textbf{Mythril}} & 
\rotatebox{90}{\textbf{Osiris}} & 
\rotatebox{90}{\textbf{Oyente}} & 
\rotatebox{90}{\textbf{Securify}} & 
\rotatebox{90}{\textbf{sFuzz}} & 
\rotatebox{90}{\textbf{Slither}} & 
\rotatebox{90}{\textbf{SmartCh}} & 
\rotatebox{90}{\textbf{solhint}}  &
\rotatebox{90}{\textbf{GPT-4o}} & 
\rotatebox{90}{\textbf{Llama-3.1-8b}}\\
\midrule
0.25 0.25,0.25,0.25 & 4.37 & 2.89 & 1.86 & 3.69 & 4.46 & 1.62 & 2.22 & 1.55 & 2.18 & 5.16 & 4.31 & 6.00 & \textbf{6.86} & 5.26 \\
0.7,0.1,0.1,0.1 & 4.42 & 3.88 & 1.51 & 3.71 & 5.28 & 2.07 & 2.30 & 1.71 & 2.51 & 4.25 & 4.07 & 4.80 & \textbf{7.49} & 5.59 \\
0.1,0.7,0.1,0.1 & 1.75 & 1.16 & 0.85 & 1.48 & 1.81 & 0.71 & 2.39 & 0.84 & 0.88 & 8.06 & 3.72 & \textbf{8.40} & 5.47 & 2.23 \\
0.1,0.1,0.7,0.1 & 4.75 & 2.35 & 3.74 & 4.47 & 4.79 & 1.25 & 1.49 & 1.82 & 1.47 & 3.86 & 4.72 & 5.40 & \textbf{5.75} & 5.10 \\
0.1,0.1,0.1,0.7 & 6.55 & 4.15 & 1.34 & 5.07 & 5.99 & 2.45 & 2.69 & 1.82 & 3.87 & 4.46 & 4.72 & 5.40 & \textbf{8.75} & 8.10 \\
\bottomrule
\end{tabular}
\end{table}

Table \ref{tab:test-score} presents the scores for each tool under five different weight configurations, emphasizing different aspects of tool performance. GPT-4o consistently achieved the highest scores across most weight configurations, indicating its well-rounded performance. Llama-3.1-8b generally performed well, often scoring second or third highest. Among traditional tools, solhint, Slither, and Mythril tended to score higher than others. When emphasis was placed on resource efficiency, solhint led with a score of 8.40, followed closely by Slither (8.06).

The choice of tool may depend on specific project priorities. For rapid, preliminary scans, tools like solhint or Slither might be preferred. For in-depth security audits, GPT-4o or a combination of tools could be most effective. The flexibility provided by different weight configurations in our scoring system allows for tool selection based on project-specific requirements.

In addition, a combined approach using multiple tools could potentially yield the most comprehensive results. For example, solhint could be used for grammatical checks and ensuring code adherence to standards. LLMs like GPT-4o or Llama-3.1-8b could identify known vulnerabilities and provide suggestions to prevent attacks. Slither could offer deeper analysis of the code and detect semantic-level issues. These findings provide clear and comprehensive responses to \textbf{RQ4} and \textbf{RQ5}, highlighting the relative strengths and weaknesses of different tool types and suggesting strategies for their effective use in smart contract vulnerability detection.

\subsection{Threat to validity}
Threats to validity are factors that have the potential to impact the results of an experiment and the validity of its findings. In our research, we have identified two specific aspects that could pose threats to the validity of our study: the categorization of smart contract vulnerabilities and the generality of the evaluation datasets.

One potential threat to the validity of our evaluation is the subjectivity and variation among researchers in evaluating and categorizing vulnerabilities and their associated smart contracts. Different researchers may have diverse perspectives, criteria, and interpretations when assessing the severity and classification of vulnerabilities. This subjectivity can introduce bias and affect the validity of the comparisons presented in the evaluation. 
To mitigate this threat, we have adopted a systematic approach based on industry standards and best practices in Section \ref{vulnerability}. We have thoroughly reviewed and discussed each vulnerability category to ensure a consistent and objective classification. This involved extensive research, consultation with experts, and careful consideration of existing literature. We have also provided clear definitions, criteria, and explanations for each vulnerability category considered in our analysis. 
By providing this transparency and documentation of our evaluation process, we aim to minimize ambiguity and facilitate a more consistent understanding of the vulnerabilities across different researchers and readers.

The generality of the evaluation datasets represents another potential threat to the validity of our research. This threat refers to the extent to which the datasets used for evaluation accurately reflect real-world scenarios and the usage patterns of smart contracts. If the evaluation datasets are limited in scope or fail to encompass the diversity of smart contract applications, the findings and conclusions may lack generalizability. 
To mitigate this threat, we have made significant efforts to address dataset limitations. We have conducted an extensive collection of contract tests from various sources, including publicly available datasets and our own developed test cases. Our dataset consists of 110 contract test cases, which have been carefully selected to cover a wide range of applications and different code sizes. 
By incorporating diverse contract test cases, we aim to provide a more representative evaluation of smart contract vulnerabilities and increase the generalizability of our findings.

While we have taken measures to address these threats, it is important to acknowledge that limitations may still exist. To further enhance the validity of future studies, researchers can focus on refining vulnerability categorization criteria and collecting larger, more diverse datasets that better capture real-world scenarios.

\section{Conclusions}
\label{discussion}

As the adoption of smart contract technology continues to surge, its security becomes increasingly critical for the robustness of blockchain ecosystems. Our comprehensive survey, encompassing vulnerabilities, attacks, defenses, and tool support, not only advances academic understanding but also offers practical implications for developers and stakeholders in blockchain technology. Our novel classifications of vulnerability types and attack patterns provide developers with a clearer understanding of potential security risks. This understanding is crucial for developing more secure smart contracts. By being aware of common vulnerabilities and how attacks exploit them, developers can proactively incorporate security measures during the development phase. Additionally, our evaluation of 14 vulnerability-detecting tools guides developers in choosing the most effective tools for their specific needs, thereby enhancing the security of their smart contracts. 
Stakeholders, including project managers and investors in blockchain projects, can leverage our findings to make informed decisions about the security aspects of the projects they are overseeing or investing in. Understanding the strengths and limitations of different security tools allows them to assess the robustness of security practices employed in their projects and advocate for the adoption of the most effective methodologies. Additionally, the annotated dataset of 110 smart contracts we created serves as a valuable resource for the community. It provides a standardized benchmark that facilitates comparative studies and ongoing evaluations of tools' effectiveness in vulnerability detection. This contributes to a collective effort in improving the overall security standards in the blockchain ecosystem.

However, the smart contract landscape is evolving rapidly, with new functionalities and protocols leading to the emergence of new security vulnerabilities. To make smart contract languages more robust, it is crucial to continue investing in research and development. For instance, there has been a growing interest in using programming languages other than Solidity for smart contract development. Languages like Go and Rust have gained attention due to their stronger syntax and logical soundness, offering potential solutions to address some of the security issues associated with Solidity. 
Furthermore, there is a need for more powerful analysis tools capable of identifying dynamic or logic errors within smart contracts. Existing tools primarily focus on known vulnerabilities and attacks, while effective methodologies for dealing with unknown attacks are still limited. Thus, protecting smart contracts from unknown attacks poses a significant challenge for future research. Additionally, developing automated approaches for repairing vulnerable smart contracts after deployment could prove to be a fruitful direction. 
In conclusion, the security of smart contracts remains an ongoing concern that demands continuous attention and innovation to address evolving threats.

\begin{acks}
  This work is supported by National Key Research and Development Program of China under the grant No.2021YFB2701202.
  This work is supported by National Natural Science Foundation of China (NSFC) under the grant No. 62172040,  No. 62002094, No. U1836212, and National Key Research and Development Program of China under the grant No.2021YFB2701200,  2022YFB2702402, and Anhui Provincial Natural Science Foundation under the grant No.2008085MF196.
\end{acks}

\bibliographystyle{ACM-Reference-Format}
\bibliography{example}

\end{document}